%% file: msastroph.tex
\documentclass[preprint]{aastex}

\newcommand{\hmpc}{h^{-1}\mbox{Mpc}}
\newcommand{\Evar}{{\cal E}}
\newcommand{\Svar}{{\cal S}}
\newcommand{\Lvar}{{\cal L}}
\newcommand{\Cvar}{{\cal C}}
\newcommand{\Kvar}{{\cal K}}

\newcommand{\ttilde}{\tilde{t}}
\newcommand{\tdyn}{t_{\rm dyn}}
\newcommand{\tcool}{t_{\rm cool}}
\newcommand{\ttdyn}{\tilde{t}_{\rm dyn}}

\newcommand{\ndotform}{\dot{N}_{\rm form}}
\newcommand{\psurv}{p_{\rm surv}}
\newcommand{\epsuv}{\epsilon_{\rm UV}}
\newcommand{\epseff}{\epsilon_{\rm eff}}
\newcommand{\epshm}{\epsilon_{\rm hm}}
\newcommand{\epscool}{\epsilon_{\rm cool}}
\newcommand{\epsstar}{\epsilon_{\ast}}
\newcommand{\epsmin}{\epsilon_{\rm min}}
\newcommand{\II}{{I\!I}}
\newcommand{\xII}{x_{\II}}
\newcommand{\fII}{f_{\II}}
\newcommand{\nhtwop}{n_{{\rm H}_2^+}}
\newcommand{\nhtwo}{n_{{\rm H}_2}}
\newcommand{\nhm}{n_{{\rm H}^-}}
\newcommand{\nhp}{n_{{\rm H}^+}}
\newcommand{\nhn}{n_{{\rm H}^0}}

\newcommand{\xhtwop}{x_{{\rm H}_2^+}}
\newcommand{\xhtwo}{x_{{\rm H}_2}}
\newcommand{\xhm}{x_{{\rm H}^-}}
\newcommand{\xhp}{x_{{\rm H}^+}}
\newcommand{\xhn}{x_{{\rm H}^0}}

\newcommand{\htwop}{{\rm H}_2^+}
\newcommand{\htwo}{{\rm H}_2}
\newcommand{\hm}{{\rm H}^-}
\newcommand{\hp}{{\rm H}^+}
\newcommand{\hn}{{\rm H}^0}

\slugcomment{Submitted to The Astrophysical Journal}

\begin{document}

\title{A Semi-Analytic Model for Cosmological 
Reheating and Reionization 
Due to the Gravitational Collapse of Structure}

\author{Weihsueh A. Chiu\altaffilmark{1}}
\affil{Joseph Henry Laboratories, Department of Physics, 
        Princeton University}
\affil{Princeton, NJ 08544}
\email{chiu@astro.princeton.edu}
\and
\author{Jeremiah P. Ostriker}
\affil{Princeton University Observatory}
\affil{Princeton, NJ 08544}
\email{jpo@astro.princeton.edu}

\altaffiltext{1}{Present address: 
Special Studies and Evaluations,
U.S. General Accounting Office,
441 G Street, NW, Washington, DC\ \ 20548}

\begin{abstract}
We present a semi-analytic model for the thermal and
ionization history of the universe at $1000 \gtrsim z \gtrsim 3$.
This model incorporates 
much of the essential physics included 
in full-scale hydrodynamical simulations, such as 
(1) gravitational collapse and virialization; 
(2) star/quasar formation and subsequent 
ionizing radiation; (3) heating and cooling; 
(4) atomic and molecular physics of hydrogen; and
(5) the feedback relationships between these 
processes.  In addition, we model 
the process of reheating and reionization using 
two separate phases, self-consistently 
calculating the filling factor of each phase. 
Thus radiative transfer is treated more accurately than
simulations to date have done: we allow to lowest order
for both the
inhomogeneity of the sources and the sinks of radiation.  
After calibrating and checking the results of this model
against a hydrodynamical simulation, we apply our model to
a variety of Gaussian and non-Gaussian 
CDM-dominated cosmological models.  
Our major conclusions include: (1) the epoch of reheating depends 
most strongly on the power spectrum of density fluctuations
at small scales; (2) because of the effects of
gas clumping, full reionization occurs at $z\sim 10$ 
in all models; (3) the CMBR polarization and the 
stars and quasars 
to baryons
ratio are strong potential discrimants between different assumed 
power spectra; (4) the formation of galactic spheroids may be 
regulated by the evolution of reheating through feedback,
so that the Jeams mass tracks the non-linear mass scale;
and (5) the evolution of the bias of luminous objects can potentially
discriminate strongly between Gaussian and non-Gaussian PDFs.
\end{abstract}


\section{Introduction}
\label{sec:intro}
In this paper we model semi-analytically the thermal and ionization 
history of the universe and its effects on the formation 
of luminous objects.  
Many parts of this complex early evolution of the universe
have been studied previously, beginning with the
papers of \citet{Binney77}, 
\citet{Silk77}, and 
\citet{RO77}, the last of whom used models 
of the thermal history of the 
universe to investigate the formation of galaxies. 
\citet{Tegetal97}  used a 
simplified model of dark matter halo profiles to derive 
the minimum mass of objects which are able to cool in a
Hubble time.  
The evolution of the intergalactic medium (IGM) 
has been explored semi-analytically 
by numerous authors, including 
\citet{SGB94}, and recently \citet{vs99}.  
Also recently, \citet{ciardi99} use a semi-analytic model for
galaxy and star formation in conjunction with high 
resolution N-body simulations.  
The detailed evolution of the IGM
was addressed in a full hydrodynamical simulation by
\citet{OG96} and \citet{GO97} 
(hereafter GO97).
These studies 
explicitly allowed for clumping of the IGM, but was only able to
simulate a small volume (a cube of side $2 \hmpc$).  
\citet{OG96} 
found that a mass resolution
of about $10^{4} M_{\odot}$ was required to resolve 
early epochs of reheating and reionization.

The more general 
problem of galaxy (as opposed to dark matter halo) formation 
has also been addressed in full 
hydrodynamical simulations by many investigators, 
including \citet{Katz96},
GO97, and recently by \citet{Virgo99}.
\citet{Virgo99} 
simulate a large volume, but with much courser
mass resolution than GO97, and thus 
was not meant to address the details of
reheating and reionization.  
\citet{Kauf97} developed a ``hybrid''
scheme, using a semi-analytic model of gas dynamics and star formation 
in conjunction with N-body simulations of the formation 
of dark matter halos.  
In most cases, including all the 
hydrodynamical studies cited here, only Gaussian CDM have 
been considered.

Here, we attempt to incorporate much of the 
essential physics modeled by GO97 into a semi-analytic 
model for the reheating and reionization of the universe.  
The work is not explicitly three dimensional.  Rather, we treat
the multi-phase medium in a statistical fashion and 
treat collapse in the approximation of the Press-Schechter 
method 
\citep{PS74}.
We consider the universe as an evolving two-phase medium --- one
phase ($I$) being neutral and the other phase ionized ($\II$).  
Of course we still must make many assumptions as to which 
physical processes are important and which may be neglected.
The main assumption we make is that UV photons 
from an early generation of stars and quasars are
the main source of energy 
for the reheating and reionization of the universe

Our model also includes the following physical processes:
(1) gravitational collapse and virialization, 
(2) non-equilibrium atomic and molecular physics of hydrogen, 
(3) heating and ionization by UV photons, 
(4) cooling, (5) simulated star/quasar formation, 
(6) clumping, and (7) feedback between star formation and
the thermal and ionization evolution of the universe.
All of these processes are calculated within the context of the
standard evolution of the background cosmological model.
In determining the evolution of structure formation, we 
explore several models for the probability density function (PDF)
and power spectrum
of the initial cosmological density field.
Once the background cosmology and normalization of the power 
spectrum are fixed, our semi-analytic model has only 
two free parameters related to the efficiency of star formation
and UV photon generation.  These two parameters we fix by calibrating
our results with GO97 and by requiring each cosmological scenario
to reproduce appoximately the observed ionizing photon intensity $z=4$.
The model we develop \emph{neglects} the following processes:
(1) spatial variations and correlations beyond the 
two phases, (2) non-virialization shock heating from 
gravitational collapse and supernova explosions, 
(3) local optical depths to UV photons and the 
frequency evolution of the photon energy spectrum, and
(4) helium, dust, and heavy elements in the IGM\null.
Because these effects are likely to be important at
late times, we only consider our models a reasonable approximation
for $z\geq 3$.

A word on motivation may be useful here.  
In almost all respects the full numerical hydrodynamical simulations
provide a better physical treatment than we are implementing here; 
for example, the neglected processes 
(1)--(4) not treated by the current work
\emph{are} included in GO97.  The two related ways that this
treatment is to be preferred are that it is not limited by 
numerical resolution and that it allows, to lowest order, for a  
multi-phase (ionized/neutral) IGM.  Given the overall relative
strengths of the full numerical treatment, one can ask why bother
to do anything less?  The answer is threefold: 
\begin{enumerate}
\item First, residual doubts about numerical resolution always
remain whatever tests are made, so it is important to see how robust
the conclusions of the numerical work are when resolution limits
are removed.
\item The full numerical approach is so expensive that only a very 
small number of models can be explored.  The result is that one is 
again concerned with the questions of robustness and generality of
the results to variations of the adopted model.  How sensitive 
are conclusions to the assumed nature of the
perturbations (Gaussian/non-Gaussian), amplitude of the power
spectrum, etc.  Semianalytic methods allow one to 
explore the parameter space.
\item The full numerical results, paradoxically, in so far
as they are better and are more and 
more comprehensive in physical treatment, tell one less and less!
Understanding depends on knowing which physical processes and 
initial conditions are \emph{essential} to the results, and 
which merely modify the details of the 
overall picture.  A semi-analytic treatment
allows one to add and subtract from the 
modeling to determine what matters and what does not.
\end{enumerate}

The organization of this paper is as follows.  
In \S~\ref{sec:model}, we present the elements of 
our adopted physical model for reheating and reionization.  In 
\S~\ref{sec:results}, we apply 
it to several Gaussian and non-Gaussian cosmological models.
We then place some constraints
on these models using several current
and potential observations, such as Gunn-Peterson absorption and 
the CMBR, which are directly 
affected by the ionization
history of the universe.  In \S~\ref{sec:discussion}, 
we discuss the relationship between the reheating and
ionization history of different models to the 
evolution of the luminous matter in the universe.
We summarize our results in \S~\ref{sec:summary}.
In the appendices, we delineate our treatment of the standard
physics of the universe, as well as the specific calculational
implementation (e.g., finite difference equations)
of our semi-analytic model.

\section{A Model for the Physics of Reionization}
\label{sec:model}

In this section, we describe the semi-analytic model we use 
of the physics of reionization.  We 
concentrate on the treatment of a two-phase medium with 
sources and sinks of radiation.  In particular we define our
basic differential equations, and describe how we calculate the
formation rate of bound objects, the UV production rate, and 
the clumping of cosmic gas.

\subsection{Definitions and Conservation Equations}

Since we are considering a two-phase medium, we will need to
understand the time evolution of the filling factor 
of ionized regions $f_{\II}$.  The two other independent 
quantities required are the ionization fraction in the 
ionized region $x_{\II}$ and the mean (volume-averaged) UV 
ionizing ($\geq 13.6$~eV $\equiv \epsilon_0$) intensity $\bar{J}$.
The corresponding intensity in the ionized region is
$\bar{J}_{\II} = \bar{J}/f_{\II}$.  In our notation,
$n_H$ is the total mean hydrogen density.

Considering only ionization of hydrogen, 
local particle conservation in the ionized region gives
\begin{eqnarray}
\frac{d}{dt}\left(n_H \xII \right) & = & -3\frac{\dot{a}}{a} n_H \xII   + 
	n_H (1 - \xII ) \frac{4\pi \bar{J}_{\II} \sigma_p}{\epsilon_0}
	\nonumber\\
	& & \qquad +\, \Cvar\, n_H^2 \left[\xII (1 - \xII) k_{ci} 
		- \xII ^2 \alpha_R\right]\\
	& = & -3\frac{\dot{a}}{a} n_H \xII   + 
	n_H (1 - \xII ) \frac{\bar{E} c\sigma_p}{\fII \epsilon_0}
	\nonumber\\
	& & \qquad +\, \Cvar\, n_H^2 \left[\xII (1 - \xII) k_{ci} 
		- \xII ^2 \alpha_R\right]
\label{eq:dnxdt_new}
\end{eqnarray}
where the first term on the right hand side is from
the expansion of the universe, and the second 
term from photoionization, and the third term from
collisional ionizations (rate coefficient $k_{ci}$)
and radiative recombinations (rate coefficient $\alpha_R$).  
The effective cross section to photoionization $\sigma_p$ 
is given below.  The clumping factor $\Cvar \equiv 
\langle \rho_b^2\rangle / \langle \rho_b \rangle^2$ 
takes into account the non-uniformity of the gas.  We
describe how to estimate $\Cvar$ in \S~\ref{subsec:clumping}.

For an energy spectrum of the form $E_\nu \propto \nu^{-\gamma}$,
the average UV energy density ($E = 4\pi J/c$) 
of the universe evolves as
\begin{eqnarray}
\frac{d\bar{E}}{dt} & = & \bar{S} 
	-[4 + (\gamma - 1)]\frac{\dot{a}}{a} \bar{E}
		- \fII n_H (1 - \xII ) 4 \pi \bar{J}_{\II} \sigma_e
	- \epsilon_0 n_H \xII  \frac{d\fII}{dt}\\
	& = &  \bar{S}
	-[4 + (\gamma - 1)]\frac{\dot{a}}{a} \bar{E}
		- n_H (1 - \xII ) \bar{E} c\sigma_e
	- \epsilon_0 n_H \xII  \frac{d\fII}{dt},
\label{eq:dedtraw_new}
\end{eqnarray}
where we have used the relationship between $\bar{E}$,
$\bar{E}_{\II}$, and $\bar{J}_{\II}$.  We use a value of
$\gamma=1.5$ throughout.
The integrated 
source function is given by $\bar{S}$, the Hubble expansion 
is accounted for in the second term, 
energy lost to
maintaining the ionization of the ionized region is 
given by the third term, and the last term is due to 
energy lost to ionizing new (i.e., previously neutral) material.
The effective UV energy loss cross section is $\sigma_e$.

The definitions of $\sigma_p$ and $\sigma_e$ are as follows.
If the ionizing specific intensity is $\bar{J}_\nu$, then
the ionizing intensity is
\begin{equation}
\bar{J} \equiv \int_{\nu_0}^{\nu_\ast} \bar{J}_\nu d\nu,
\end{equation}
where $\nu_\ast$ is a high frequency cutoff (we use 7.4 keV, 
from Abel 1998).
For a power law 
\begin{equation}
\bar{J}_\nu = J_0\nu_0^{-1}\left(\frac{\nu_0}{\nu}\right)^\gamma,
\end{equation}
we have the relation
\begin{equation}
\bar{J} = \frac{J_0}{\gamma - 1}\left[1 - 
	\left(\frac{\nu_0}{\nu_\ast}\right)^{\gamma-1}\right].
\end{equation}
The photoionization rate per hydrogen atom is given (in s$^{-1}$) by
\begin{eqnarray}
\Gamma_p & \equiv & \int_{\nu_0}^{\nu_\ast} 4\pi \bar{J}_\nu \sigma(\nu) 
	\frac{d\nu}{h_p\nu}\\
	& = & \frac{4\pi J_0}{h_p \nu_0}
	\int_{\nu_0}^{\nu_\ast}\left(\frac{\nu_0}{\nu}\right)^\gamma
		\sigma(\nu)\frac{d\nu}{\nu},
\end{eqnarray}
where $h_p$ is Planck's constant.  Equating this with 
$4\pi\bar{J}\sigma_p/\epsilon_0$ gives
\begin{equation}
\sigma_p \equiv 
	\frac{\gamma - 1}{1 - \left(\frac{\nu_0}{\nu_\ast}\right)^{\gamma-1}}
	\int_{\nu_0}^{\nu_\ast}\left(\frac{\nu_0}{\nu}\right)^\gamma
		\sigma(\nu)\frac{d\nu}{\nu}.
\end{equation}

Similarly, assuming that all the excess energy
from each photoionization goes into heat and produces no UV photons, 
the rate of UV energy loss per hydrogen atom (erg~s$^{-1}$) is 
\begin{eqnarray}
\Gamma_e & \equiv & \int_{\nu_0}^{\nu_\ast} 4\pi \bar{J}_\nu \sigma(\nu)d\nu\\
	& = & 4\pi J_0
	\int_{\nu_0}^{\nu_\ast}\left(\frac{\nu_0}{\nu}\right)^\gamma
		\sigma(\nu)\frac{d\nu}{\nu_0}.
\end{eqnarray}
Equating this with $4\pi\bar{J}\sigma_e$ gives
\begin{equation}
\sigma_e \equiv
	\frac{\gamma - 1}{1 - \left(\frac{\nu_0}{\nu_\ast}\right)^{\gamma-1}}
	\int_{\nu_0}^{\nu_\ast}\left(\frac{\nu_0}{\nu}\right)^\gamma
		\sigma(\nu)\frac{d\nu}{\nu_0}.
\end{equation}

The difference between these cross sections
gives the cross section for photoheating, or equivalently a 
photoheating rate (erg~s$^{-1}$) of
\begin{equation}
\Gamma_{ph} \equiv 4\pi\bar{J}(\sigma_e - \sigma_p).
\label{eq:gamma_ph}
\end{equation}

In order to fully specify the model, we need to be able to determine
$\bar{S}$, as well as have a relationship between $\bar{E}$ and
$\fII$.  These turn out to be intimately related.

\subsection{Formation Rate of Bound Objects}

We assume that stars/quasars only form in virialized collapsed objects.  
Hence the first input into the star formation rate is the formation
rate of bound objects.  In the next subsection we 
relate these bound
objects to UV energy output in order to derive expressions for the
source function $\bar{S}$ and the relationship between the
mean radiation energy density $\bar{E}$ and the filling factor
$\fII$.  

Generalized Press-Schechter for a PDF $P(\nu)$, 
and normalization factor $f^{-1}\equiv \int_0^\infty P(\nu)d\nu$
gives the comoving number density of collapsed objects with initial 
comoving radius between $R$ and $R+dR$ to be
\begin{eqnarray}
N_R dR & = & f\cdot P\left[\nu(R,t)\right] 
	\frac{3}{4\pi R^3} \frac{\partial\ln\nu(R,t)}{\partial\ln R} 
	\nu(R,t) \frac{dR}{R}
\label{eq:ps_nrdef}
\\
\nu(R,t) & \equiv & \frac{\delta_c}{\sigma(R)\,D(t)}
\label{eq:ps_nudef}
\\
\frac{\partial\ln\nu(R,t)}{\partial\ln R} & = &
	\frac{d\ln\sigma^{-1}(R)}{d\ln R} \equiv \alpha(R)
\label{eq:ps_alphadef}
\end{eqnarray}
where $\sigma(R)$ is the rms fluctuation on scale $R$ linearly
extrapolated to the present, $\delta_c \approx 1.69$ is the 
standard linear overdensity of collapse, and
$D(t)$ is the linear growth factor normalized to 
$D(t_0)=1$.  The original Press-Schechter \citep{PS74}
method was derived in for a Gaussian PDF, and has been
checked in that case with N-body simulations \citep{Efst88}.
These expressions in the non-Gaussian case 
were derived in \citet{cos97},
and recently tested through N-body simulations 
for several non-Gaussian models 
\citep{RB99}.

For each model, we filter the 
power spectrum using a top-hat window 
to obtain the rms fluctuation $\sigma$.  
The power spectra we use are from 
\citet{BBKS} and 
\citet{Sugiyama95} in the adiabatic cases; 
\citet{pen94} and 
\citet{pen95} in the textures case;
and Peebles 1999ab in the ICDM case.

In general, we cannot use only the Press-Schechter
number density to determine the formation rate of bound objects. 
This is because the net time-rate of change of the number 
density of objects at a particular scale consists of both a 
formation $\dot{N}_{R,\rm form}$ rate and a
destruction rate $\dot{N}_{R,\rm dest}$.  Thus the actual formation
rate may be larger than the rate inferred from Press-Schecter alone.
Of course, we must constrain these rates to be consistent
with the Press-Schechter formalism:
\begin{equation}
\dot{N}_R = \dot{N}_{R,\rm form} - \dot{N}_{R,\rm dest},
\label{eq:psconsistency}
\end{equation}
where $\dot{N}_R$ is given by differentiating 
equations~(\ref{eq:ps_nrdef})--(\ref{eq:ps_alphadef})
with respect to time.
We seek a semi-analytic methodology 
applicable to an arbitrary PDF which gives these two rates. 

\citet{Sasaki94}
models the formation and destruction 
rates of these objects by assuming that the destruction efficiency
rate $\phi \equiv \dot{N}_{R,\rm dest}/N_R$ 
has no characteristic scale --- ``destruction''
by incorporation into larger objects is hierarchical and
scale-invariant.  In this section, we generalize his derivation
to non-Gaussian PDFs.

The assumption of no characteristic scale of destruction efficiency, 
in addition to the requirement 
that the net rate of formation minus destruction be 
consistent with Press-Schechter,
leads to the conclusion that $\phi$ must 
be a function only of time.  \citet{Sasaki94}
derives
the following relation for the destruction probability per
unit time:
\begin{equation}
\phi(t) = \frac{\dot{D}}{D},
\end{equation}
which we find to be independent of the PDF.  It follows that 
the probability that an object created at $t_1$ 
exists at $t_2$ without having been 
incorporated into a larger object 
is given by
\begin{equation}
p_{\rm surv}(t_2|t_1) = \exp\left[\int_{t_1}^{t_2} \phi(t)dt\right]
 	= \frac{D(t_1)}{D(t_2)}.
\label{eq:psurv_def}
\end{equation}

Combining $\phi$ and the Press-Schechter consistency requirement
equation~(\ref{eq:psconsistency})
leads to a formation rate of the form:
\begin{equation}
\dot{N}_{R,\rm form}  =  N_R \cdot 
	\nu(R,t) \frac{\dot{D}}{D}\left(\frac{-P'}{P}\right),
\end{equation}
where $P' = dP/d\nu$.
The quantity $-P'/P$ for a Gaussian
and an exponential $P(\nu)$ takes on the forms:
\begin{eqnarray}
\frac{-P'}{P} & = & \nu,\qquad \mbox{if}\ 
	P(\nu)\propto e^{-\frac{1}{2}\nu^2};
\label{eq:pprime_p_gaus}
\\
\frac{-P'}{P} & = & \omega,\qquad \mbox{if}\ P(\nu)\propto e^{-\omega\nu}.
\label{eq:pprime_p_exp}
\end{eqnarray}
Textures and non-Gaussian ICDM both have PDFs which 
are well approximated by exponentials 
(Gooding et al. 1992; Park, Spergel, \& Turok 1991; 
Peebles 1998ab).
For textures $\omega = 1.45$ (using Park, Spergel, \& Turok 1991 
as described in
Chiu, Ostriker, \& Strauss 1997), 
and for non-Gaussian ICDM $\omega = 0.67$ \citet{Peebles99b}.
Recalling that 
\begin{equation}
	\frac{\dot{D}}{D} = f(\Omega)\frac{\dot{a}}{a},
\end{equation}
where the velocity factor $f(\Omega)\approx \Omega^{0.6}$,
the formation rate in closed form is 
\begin{equation}
\dot{N}_{R,\rm form} dR  =  f\cdot 
	\left[-P'(\nu)\right]
	\frac{3}{4\pi R^3}
	f(\Omega)\frac{\dot{a}}{a}
	\alpha(R)\,
	\nu^2(R,t)\, \frac{dR}{R} .
\end{equation}
This is the formation rate per comoving volume we use in our model. 
For a power law $\sigma(R)\propto R^{-\alpha}$, 
the logarithmic slope 
$\partial\ln\nu/\partial\ln R = \alpha$ is a constant.

This formalism for merging and formation rates is consistent
with the work of
\citet{BL93}, who used a numerical extension 
of Press-Schechter with simple forms for the merging probability.
All other work of this type is based on \citet{LC93} and 
\citet{LC94},
who construct conditional probabilities by
extending the work of \citet{Bondetal91}.
Unfortunately, this 
method is heavily dependent upon the PDF begin Gaussian, and is 
sufficiently complicated so that extending it to non-Gaussian 
PDF is not straight forward.  Furthermore, since objects are
``constantly growing'' in the 
\citet{LC93} formalism, 
rather than in ``punctuated equilibrium'' as implied by 
equation~(\ref{eq:psurv_def}), it is 
difficult to directly compare the two methods.  However, one
sign that they might be roughly consistent is that 
\citet{VL96} obtain similar results for cluster abundances 
whether they use the 
\citet{LC93} or the \citet{Sasaki94} 
methodology.  

We further note that during reheating and reionization,
structure formation is largely hierarchical in most of the models
we consider.  This is because only a small fraction of the 
universe needs to collapse in order to release enough energy in 
order to reheat and reionize the universe.  Collapsed objects
are thus still ``rare'' ($\nu > 1$), and their net formation rates
are thus dominated by their creation rates.  The use of the
Sasaki results thus provide only a small correction.  
In the limit where the ``destruction'' rates are negligible, 
all semi-analytic merging formalisms which are based on
Press-Schechter must give the same answer in order to be 
consistent with equation~(\ref{eq:psconsistency}). 

\subsection{Determining the UV Production Rate}

Our basic model for UV production is that individual halos
which can undergo star formation will emit UV radiation 
radially, and thus be surrounded by a sphere of ionized material, 
a ``cosmological Str\"omgren sphere.''
Consider halos of baryonic mass in the range $M_b$ to $M_b+dM_b$,
corresponding to comoving length scale between $R$ and $R+dR$, 
formed in the time interval $\ttilde$ to $\ttilde + d\ttilde$, 
and ``observed'' at time $t$.  The comoving number density
of such objects is given by
\begin{equation}
N(R,\ttilde,t)\,dR\,d\ttilde = 
	\ndotform(R,\ttilde)\cdot\psurv(t|\ttilde)\,dR\,d\ttilde,
\label{eq:n_r_tt_t}
\end{equation}
where the notation is the same as in the previous section.  
The second factor
accounts for the destruction objects of scale $R$ through merging 
into larger objects.  Although we are using Sasaki's method
to determine $\ndotform$ and $\psurv$ as described above, 
the form of equation (\ref{eq:n_r_tt_t}) is completely
general.  Note that $M_b$ and $R$ are related by
$M_b = 4\pi R^3 \rho_{b,0}/3$.  
Following the prescriptions of 
\citet{Cen92} 
and \citet{Gnedin96},
we assume that star formation is spread over a dynamical
time scale, so that the UV luminosity of such an object is given by
\begin{equation}
\ell(M_b,\ttilde,t) = 
	(1 - f_\ast)
	M_b c^2\frac{\epscool \epsstar \epshm \epsuv}{\ttdyn}
	\left(\frac{t-\ttilde}{\ttdyn}\right)
	\exp\left[-\left(\frac{t-\ttilde}{\ttdyn}\right)\right],
\label{eq:lumeq}
\end{equation}
where $f_\ast$ is the fraction of the baryonic mass which has
already been processed into stars/quasars and 
$\ttdyn$ is the dynamical time associated with the 
object at its time of formation (see equation \ref{eq:tdyn}).  
We are also assuming here that
the stars/quasars are uniformly distributed throughout the collapsed
halo.

The efficiencies $\epsilon_i$ are as follows: 
$\epscool$ is the mass fraction of the halo which is cooling, 
$\epsstar$ is the mass fraction of cooling gas which physically ends  
up in collapsed objects,
$\epshm$ is the mass fraction of collapsed objects which are
quasars or high-mass stars (i.e., with substantial UV flux), and
$\epsuv$ is the mass-to-UV efficiency of the high-mass stars/quasars. 
The cooling fraction $\epscool$ we derive below in 
Appendix~\ref{sec:gascooling_virial}, and
is a function of the halo virial mass and the redshift of collapse.  
GO97 uses
$\epsuv=6\times 10^{-5}$ and \citet{Gnedin96}
gives $\epshm = 0.16$,
derived using a standard initial mass function.
We cannot use $\epsuv$ directly because we are not 
resolving the mass elements within individual cooling halos.
The ``resolution'' factor $\epsstar$ parameterizes this
uncertainty.  We use a net efficiency 
\begin{equation}
\epseff \equiv \epsstar \epshm \epsuv,
\end{equation}
normalized by requiring, for each model, that the ionizing 
intensity $J_{21} = 1$ at $z = 4$.  There is, effectively, only
\emph{one} free parameter here, $\epseff$, and that is fixed by 
normalizing to observations.  

A separate estimate of $\epsstar$, however, is necessary in order to 
derive $f_\ast$, the fraction of baryons which is sequestered
in stars/quasars and their remnants.  It is important to keep 
track of $f_\ast$ to avoid ``double counting'' star formation 
--- baryons which are in such bound objects are not available for
additional star formation.  Note that 
results are insensitive to the value
of $f_\ast$ while $f_\ast\lesssim 10\%$, 
since they are only dependent on $1-f_\ast$.
We can use the GO97 simulation
to derive our ``resolution'' factor $\epsstar$ by assuming
that it is constant for all cosmological models.  If $\epsstar$
is a constant, then it should correspond to the value
which gives the same $J_{21}$ as the GO97
simulation, using their values for $\epshm$ and $\epsuv$ and
the same cosmological model.
This correspondence leads to a value $\epsstar \approx 0.5$.

Given $\epsstar$, the simplest way to calculate the fraction $f_\ast$ is
actually to integrate over the 
UV production rate $\bar{S}$, given in equation (\ref{eq:sbar_int_l}), 
dividing by the appropriate efficiencies:
\begin{eqnarray}
f_\ast(t) & = & \int_0^t \frac{\bar{S}(\ttilde) d\ttilde}
{\epshm \epsuv n_H m_H c^2/X}
\\
& = &\epsstar \int_0^t \frac{\bar{S}(\ttilde) d\ttilde}
{\epseff n_H m_H c^2/X},
\label{eq:fstar_int_eq}
\end{eqnarray}
where $X$ is the hydrogen mass fraction, and we have
expressed the result in terms of the $J_{21}$-normalized
effective efficiency $\epseff$ and the resolution factor
$\epsstar$.
As a check on our value of $\epsstar$, 
we compare $f_\ast$ at $z \approx 4$ between our model and the 
GO97 simulation, given the same cosmological
parameters.  We find the values to be consistent, 
with $f_\ast \approx 2\%$ at this redshift (see also 
Figure~\ref{fig:go_comp_twoplot}).

The global mean UV production rate per unit (physical) volume 
is given by
\begin{equation}
\bar{S}(t)  =  a(t)^{-3}\int_0^{\infty} dR \int_0^t d\ttilde\,
	\psi(R,\ttilde) \cdot
	N(R,\ttilde,t)\cdot \ell(M_b,\ttilde,t),
\label{eq:sbar_int_l}
\end{equation}
where the function $\psi(R,\ttilde)$ is a selection function for
objects which are able to collapse and form stars/quasars.  
One simple form for $\psi$ is 
\begin{equation}
\psi_{\rm simp}(R,\ttilde) = \theta[R - R_J(\ttilde)],
\end{equation}
where the $\theta$ function is zero if the argument is $<0$ and
unity if the argument is $>0$. 
These represent the criteria
that a mass must be greater than the Jeans mass to collapse.  
In order to account for the fact that 
the Jeans mass will be different in the two phases, we
can define $\psi$ to be
\begin{equation}
\psi_{\rm 2ph}(R,\ttilde) = \left\{ 
	(1-\tilde{f}_{\II})\cdot \theta\left[R - R_{J,c}(\ttilde)\right] + 
	\tilde{f}_{\II}\cdot \theta\left[R - R_{J,h}(\ttilde)\right]\right\},
\label{eq:twophase_select_func_simp}
\end{equation}
where $\tilde{f}_{\II}$ is the filling factor of the ionized region
at time $\ttilde$,
$M_{J,c}$ is the Jeans mass in the neutral (cold) region, 
and $M_{J,h}$ is the Jeans mass in the ionized (hot) region. 
Note that we are assuming that an object, once formed, can only
be destroyed by merging and forming a larger object.  We are
assuming, thus, that virialized gas halos are ``self-shielded'' 
from ionizing radiation due to their high density.  That is,
they can not be heated to the point of dynamical instability 
nor can they be ``evaporated'' away.

In order to derive the needed relation between $\fII (t)$ and 
$\bar{E}(t)$, we make two further assumptions.  
One is that the ionized volume $v_{\II}(M_b,\ttilde,t)$ 
around an isolated object is proportional to its UV
luminosity $\ell(M_b,\ttilde,t)$ with a 
coefficient of proportionality $\kappa(t)$ that is a function \emph{only}
of the time $t$ when it is ``observed.''  This should be a reasonable
approximation, since the recombination rates are determined by
the density and temperature at $t$, not at the formation time 
$\ttilde$.  As long as objects live longer than the recombination time,
our assumption is consistent with the work of 
\citet{SG87} on
cosmological H~II regions.
We thus have the relation
\begin{equation}
\ell(M_b,\ttilde,t) = \kappa^{-1}(t)\,v_{\II}(M_b,\ttilde,t) \equiv 
	\frac{4\pi}{3\kappa(t)}r_{\II}^3.
\label{eq:l_kappa_v}
\end{equation}

The second assumption, which is needed to determine
$\kappa(t)$ from $\fII (t)$ and $\bar{S}(t)$, 
is that the filling factor can be related
to the average number of ``overlapping'' ionized regions by 
modeling it as an uncorrelated Poisson process.  
The average number $\lambda(t)$ of overlapping regions  
seen by a random volume element is 
\begin{equation}
\lambda(t) \equiv 
	\int_0^{\infty} dR \int_0^t d\ttilde\,
	\psi(R,\ttilde) \cdot
	N(R,\ttilde,t)\cdot v_{\II}(M_b,\ttilde,t).
\label{eq:lambda_q_int_v}
\end{equation}
We no longer have the factor $a^{-3}$ because $v_{\II}$ is 
a comoving volume.
For Poisson probabilities, the filling factor $\fII (t)$ is
related to $\lambda(t)$ by
\begin{equation}
\fII (t) = 1 - e^{-\lambda(t)}.
\end{equation}
Combining equation (\ref{eq:lambda_q_int_v}) with 
equation (\ref{eq:l_kappa_v}) and using (\ref{eq:sbar_int_l}) 
gives
\begin{equation}
\lambda(t) = \kappa(t)\,a^3(t)\,\bar{S}(t).
\label{eq:lambda_kappa_sbar}
\end{equation}
Thus we have determined $\kappa(t)$ in terms of $\fII (t)$ and
$\bar{S}(t)$.  

With $\kappa(t)$ defined, we can derive
a relationship between $\fII (t)$ and $\bar{E}(t)$ which will 
close our system of equations.
Consider the 
contribution of each source to the mean intensity $\bar{J}(t)$.
Neglecting the optical depth within an isolated ionized region,
and assuming no flux escapes beyond a radius $r_{\II}(M_b,\ttilde,t)$,
the average flux seen by a random volume element 
due to an object of luminosity $\ell(M_b,\ttilde,t)$ is
\begin{eqnarray}
d[4\pi \bar{J}(t)] & = & 
	dR\, d\ttilde\, N(R,\ttilde,t)\int_0^{r_{\II}} d^3r
	\frac{\ell(M_b,\ttilde,t)}{4\pi[a(t)\, r]^2} \\
	& = & dR\, d\ttilde\,N(R,\ttilde,t)\cdot \ell(M_b,\ttilde,t)\cdot
	r_{\II}(M_b,\ttilde,t)\cdot a^{-2}(t)\\
	& = & dR\, d\ttilde\,N(R,\ttilde,t)\cdot \ell^{4/3} (M_b,\ttilde,t)
	\left[\frac{3\kappa(t)}{4\pi}\right]^{1/3} a^{-2}(t),
\label{eq:d4pij_int}
\end{eqnarray}
where $r$ is a comoving length, $a\,r$ is a physical length, 
and we have used
\begin{equation}
r_{\II}(M_b,\ttilde,t) = 
	\left[\frac{3 \kappa(t)}{4\pi} \ell(M_b,\ttilde,t)\right]^{1/3},
\end{equation}
which we assume to be much smaller than the Hubble radius.
Equation~(\ref{eq:d4pij_int}) also assumes that the central luminous
object emits radiation radially, which is a good approximation as
long as the radius of the sphere is much larger than the virial radius. 
Using equation (\ref{eq:lambda_kappa_sbar}) in equation 
(\ref{eq:d4pij_int}) gives
\begin{equation}
d(4\pi \bar{J}) =  dR\, d\ttilde\,
	N(R,\ttilde,t)\cdot \ell^{4/3}(M_b,\ttilde,t)
	\left[\frac{3\lambda(t)}{4\pi\bar{S}(t)}\right]^{1/3} a^{-3}(t).
\end{equation}
Therefore the mean energy density $\bar{E}(t) =  4\pi \bar{J}(t)/c$ at
time $t$ is given by
\begin{equation}
\bar{E}(t) = \frac{1}{c}
	\left[\frac{3\lambda(t)}{4\pi\bar{S}(t)}\right]^{1/3} a^{-3}
	\int_0^\infty dR \int_0^t  d\ttilde\,
	\psi(R,\ttilde)\cdot N(R,\ttilde,t)\cdot \ell^{4/3}(M_b,\ttilde,t).
\label{eq:ebar_lambda_rel}
\end{equation}
This is the desired relationship between $\fII (t) = (1-e^{-\lambda(t)})$ 
and $\bar{E}(t)$ which closes the system of equations.

We have neglected spatial correlations among these 
sources.  This approximation should be less important for 
the texture model, which is well approximated by 
uncorrelated seeds, than for the Gaussian models.  
As long as the filling factor is
small, though, these correlations are likely not to be important, since
most of the universe is neutral.  
We will find below that in the Gaussian cases we consider, 
reionization occurs quite rapidly, and this rapid percolation 
likely overwhelms any spatial correlations, which tend to slow 
down the percolation. 

Furthermore, we have neglected the optical depths within the 
collapsed object as well as in the 
gas in the Str\"omgren sphere surrounding each object.  We assume
both that the covering factor of other halos within 
the sphere is small, and that there is a sharp transition 
between the ionized and neutral regions.  
This approximation is somewhat alleviated by assuming 
a clumping factor for the gas (see \S~\ref{subsec:clumping}).  
With clumping, our model is self-consistent in the volume
average, accounting for all photoionizations and recombinations
in the ionized region.

In summary, we assume that star formation in collapsed 
halos is the source of UV photons.  The number density
of these sources is determined using the Press-Schechter 
formalism with the requirement that the baryonic halo mass is
greater than the Jeans mass; the luminosity of each source
is determined by the fraction of cooling gas and several
efficiency factors.  These sources are assumed to be 
scattered randomly (i.e., a Poisson point process), and each
surrounded by a Str\"omgren sphere.  The volume emmissivity of
these objects is assumed to be uniform, and they are assumed to
radiate radially into the surrounding intergalactic medium. 
Assuming that at fixed time, 
the volume of each Str\"omgren sphere is proportional 
to the luminosity of the central source, 
the radius of each sphere is determined by 
requiring consistency among the 
filling factor, the mean intensity, the number density of
sources, and the total UV luminosity.

\subsection{Nonuniformity of Cosmic Gas: The Clumping Factor
and Tracking the Diffuse Component}
\label{subsec:clumping}

The importance of clumping in determining the ionization history
of the universe was stressed by GO97.  
Because the clumping factor has been neglected in most
previous semi-analytic work on reionization, there has been no attempt
to estimate it analytically.  In this subsection we use 
Press-Schechter to estimate the clumping factor.  A similar procedure
was described in \citet{vs99}.

Consider dividing the  baryons in the universe into clumped and diffuse 
components.  
The clumped component consists of baryons
which have collapsed into virialized halos.  
Using equation (\ref{eq:n_r_tt_t}), 
the fraction of the baryonic mass of the universe 
which collapsed $df_m$ in
time interval $\ttilde$ to $\ttilde+d\ttilde$, 
and survives without being incorporated into
a larger object to time $t$, 
is given by
\begin{equation}
df_m(\ttilde) \equiv 
	\left[\int dR\,\frac{D(\ttilde)}{D(t)} \dot{N}_{\rm form}(R,\ttilde) 
	\frac{4\pi R^3}{3}\, \psi(R,\ttilde)\right] \, d\ttilde,
\end{equation}
where $\psi(R,\ttilde)$ is the selection function which 
filters out objects below the Jeans scale.  
The fraction 
of baryonic mass which is in collapsed objects 
at time $t$ is given by integrating over $\ttilde$:
\begin{equation}
f_m(t) = \int^t \frac{df_m(\ttilde)}{d\ttilde} d\ttilde.
\end{equation}
Note that we do not take the collapsed fraction $f_m(t)$ to be
equal to be the fraction of mass above the Jeans mass at
time $t$.  This is because during reheating, 
the Jeans mass \emph{increases},
but objects below the Jeans mass which formed at earlier
times (when the Jeans mass was lower) can still survive 
as long as they are not merged into larger objects.  
If all of these objects are in virialized halos, then
their overdensity relative to the mean density will by
\begin{equation}
\frac{\rho_{\rm vir}}{\bar{\rho}}
	\equiv \bar{\delta}_{\rm vir} \equiv 
	\frac{\Delta_{\rm vir}}{\Omega(t)},
\end{equation}
where $\Delta_{\rm vir}$ is the overdensity relative to
the critical density.  Thus the fraction of the
volume in the universe that they occupy will be
\begin{equation}
f_v(t) = \frac{f_m}{\bar{\delta}_{\rm vir}}.
\end{equation}
Self-consistency requires that the diffuse component be
underdense so that $\rho_{\rm diff} < \bar{\rho}$.  
Taking volume average gives
\begin{equation}
\bar{\rho} = f_v \rho_{\rm vir} + (1 - f_v) \rho_{\rm diff},
\end{equation}
which implies that
\begin{equation}
\frac{\rho_{\rm diff}}{\bar{\rho}} = \frac{1 - f_m}{1 - f_v}.
\label{eq:perc_rho_diff}
\end{equation}

The volume-averaged square density will be
\begin{eqnarray}
\langle \rho^2\rangle & = & f_v\rho_{\rm vir}^2 + (1 - f_v)\rho_{\rm diff}^2\\
	& = & \bar{\rho}^2 \left( f_m \bar{\delta}_{\rm vir}
		+ \frac{(1 - f_m)^2}{1 - f_v} \right).
\end{eqnarray}
The average clumping factor $\Cvar$ is thus
\begin{eqnarray}
\Cvar & \equiv & \langle \rho^2 \rangle/\bar{\rho}^2 \\
	& = & f_m \bar{\delta}_{\rm vir}
		+ \frac{(1 - f_m)^2}{1 - f_v}.
\label{eq:clumping}
\end{eqnarray}
This is the clumping factor we use in our model.  
Our model, however, assumes that UV sources are uniformly 
distributed throughout collapsed halos and that all halos 
carry the same overdensity.  These two assumptions will tend to 
reduce clumping, so equation (\ref{eq:clumping}) probably gives 
a \emph{lower} bound on the importance of clumping.  
As shown in Figure~\ref{fig:go_comp_twoplot}, 
the results of this semi-analytic calculation are in fair 
agreement with
the results of the numerical simulation in 
GO97. 

We keep track of the ionization and temperature
history of the purely diffuse phase of the ionized region.  We do this
in order to properly calculate the Gunn-Peterson optical depth, which
probes the diffuse intergalactic medium (IGM).
This phase is by definition unclumped,
and is underdense
relative to the mean, equation~(\ref{eq:perc_rho_diff}).  Note, however,
that this diffuse phase plays no feedback role --- it simply responds
to the other physics which is going on.

\section{Results}
\label{sec:results}

In this section we present the results of our calculations
for a range of cosmological models.  We begin with a ``calibration''
model which uses the same cosmological parameters as were
used by GO97 ($\Omega_0=0.35$, $\Omega_\Lambda=0.65$, $h=0.70$,
and $\Omega_b=0.03$).
We then consider 
adiabatic Gaussian CDM 
(G+ACDM), textures (T+CDM), and non-Gaussian isocurvature
CDM (ICDM) as proposed by Peebles (1999ab).
To help assess the 
sensitivity of the results to the PDF and the power spectrum,
we also consider an ``artificial'' case of a texture PDF with 
an adiabatic CDM power spectrum (T+ACDM).  
We use a Hubble parameter $h=0.65$, a baryon
density parameter $\Omega_b h^2=0.0125$ throughout, and
consider flat ($\Lambda$-dominated) and open models 
with matter density parameters of $\Omega_0=$~1, 0.45, 0.35, and 0.25.
All models are normalized to cluster abundances 
(method of Chiu, Ostriker,
\& Strauss 1997 for all models except ICDM, which
uses Peebles 1999ab).  
The cosmological parameters associated with
the models we consider are listed in the
first three columns of Table \ref{tab:results}.
As noted above, the UV efficiency
$\epseff$ in each run is adjusted to reproduce $J_{21}=1$ at
$z=4$.  There are no other adjustable parameters in the model.

\subsection{Comparison with GO97 Simulation}

In Figure~\ref{fig:go_comp_fourplot} we show the filling factor 
$f_{\II}$, the ionizing intensity $J_{21}$, the neutral fraction
$\langle 1-x\rangle$, and the gas temperature $T$ in the case were
we mimic the GO97 calculation.  While the ionizing intensity
was adjusted to be consistent with the GO97 value at $z=4$, 
the evolution of $J_{21}$ is similar in the two calculations.  
As in the GO97 calculation, 
reionization to $1-x = 10^{-3}$ occurs suddenly at 
$z \sim 10$, while reheating occurs more gradually.
The main difference between
this work and GO97 is that reionization and reheating begin 
somewhat earlier in our model.  This may in part be due to the 
limited resolution and single-phase treatment of GO97.

In Figure~\ref{fig:go_comp_twoplot}, we show the stellar fraction and
clumping factor for this reionization calculation.  Again,
the values are similar to those in GO97.  The clumping
factor shows the greatest discrepancy, and is without doubt inaccurately
calculated in our model.  Without full-scale simulation, however,
it is difficult to calculate the clumping factor to a much greater 
accuracy than what we have presented in \S~\ref{subsec:clumping}.
The fact that our result is within a factor of a few of the 
GO97 calculation is encouraging.  The earlier onset of star formation
shown in the top panel by our semi-analytic model may reflect the
limited resolution of GO97, which while adequate for
average objects, is insufficient for early, dense small ones.

In Figure~\ref{fig:go_comp_mj_mnl}, we show the nonlinear and 
average Jeans mass scales derived from our model.  Our
results are qualitatively 
consistent with
OG96: \emph{we find that the Jeans mass and the nonlinear mass (here 
defined by $\delta M/M \approx 0.4$) track each other to remarkable
accuracy during the reheating phase.}  Because reheating begins
earlier in this model than in the GO97 simulation, the precise 
degree of nonlinearity which tracks the Jeans mass is different, 
and the epochs 
during which the Jeans and nonlinear mass scales track each other
occur earlier.  A possible physical basis for this tracking is clear.
When the Jeans mass is lower than the nonlinear mass, star formation
accelerates so that the average temperature increases, the Jeans mass
increases to above the nonlinear mass scale, and further star formation
is quenched.

We do not expect an exact correspondence between our method 
and the GO97 work.  For example, 
this semi-analytic calculation is not limited by mass resolution,
so mass condensations down to the Jeans mass limit can contribute to
the ionizing UV flux.  This may partially explain why reheating
starts at $z\sim 30$ in our model, rather than at $z\sim 20$ in
GO97.  Another contributing factor to this difference in 
the reheating epoch is that the largest GO97 simulation lacked 
significant large-scale power relative to the true power spectrum.
Furthermore, we model here a two-phase 
medium where the radiation field is not spatially uniform, while
GO97 use an average UV radiation field. 
On the other hand, this calculation neglects all spatial correlations,
which are automatically taken into account by the three-dimensional
numerical calculation of GO97, and uses a power-law radiation spectrum.  
While it is unclear what the net 
effect of all these differences should be, it is encouraging that
two completely different methods of calculating the reheating
and reionization of the universe give such similar results.

\subsection{Reheating and Reionization in Various Cosmologies}

The results of our calculations for the models we consider 
are presented in Table~\ref{tab:results} and 
Figures~\ref{fig:gaus_flat_fourplot}--\ref{fig:icdm_flat_fourplot}.
Qualitatively, the evolution of reheating and reionization can be
divided into two broad categories characterized by late or early 
reheating, corresponding to late or early 
evolution of the filling factor $f_{\II}$.  In models with
late reheating, which include the Gaussian models and the 
T+ACDM model, 
the filling factor rises from $f_{\II}<10^{-4}$ at 
$z=30\sim50$ to $f_{\II}\sim 1$ at $z= 10\sim 20$.  For models with
early reheating, which include the pure texture T+CDM models and the
ICDM model, the filling factor is already $f_{\II}>0.1$ at 
$z\sim 80$, but rises slowly to $f_{\II} = 1$ at $z= 10\sim 20$.

The general picture of the evolution in these models is as follows.
We note that by ``reionization'' we mean when $f_{\II}$ is approaching
unity, but that by ``full reionization'' we mean the epoch at which
the individual ionized regions start overlapping substantially, so that 
$\lambda = -\ln{(1 - f_{\II})} \gg 1$.  

For an adiabatic CDM-type power spectrum, the fluctuation spectrum
$\sigma_R$ rises logarithmically, or nearly logarithmically, as
$R\rightarrow 0$.  Hence, at the Jeans scale of $\sim 0.01\hmpc$, 
there is a logarithmic cutoff in power for $z > \sigma_{R_J} \sim 10-20$.  
Furthermore, cooling becomes less efficient at the 
Jeans scale at later times.  These two factors
lead to a suppression of reheating and reionization until late
times.  Full reionization occurs after the filling factor
approaches unity, at redshifts $z \lesssim 10$.

These results for Gaussian CDM models 
are consistent with the work of GO97 (who considered a 
flat $\Omega_\Lambda=0.35$ CDM model), 
as well as \citet{ciardi99} (who considered
an $\Omega_0=1$ CDM model), 
but vary somewhat from the results of 
\citet{vs99} (who considered $\Omega_0=1$ and open $\Omega_0=0.3$
CDM models).  The latter find a slightly lower
value for the redshift of reionization ($z\sim 6-7$).  
This difference is unlikely due to gas clumping, 
as proposed by \citet{ciardi99}, since
both this work and GO97 incorporate clumping.  In fact, 
our work uses a
prescription for clumping very similar to \citet{vs99}.  
We do, however, use a different prescription for calculating the 
radius of the cosmological Str\"omgren spheres, but the 
rapidity with which reionization occurs may mitigate the effect of 
this differences.  While the exact value of the epoch of reionization 
remains uncertain, various methods, including ours, 
give the range $z=8\pm 2$ for Gaussian CDM models.  

For T+CDM and ICDM power spectra, the fluctuation spectrum
rises as a power law for small $R$, the Jeans scale becoming
nonlinear with $\delta M/M \sim 1$ at $z\sim 50$.  
At this redshift, cooling at the Jeans scale is still 
efficient, and UV photons are created in
abundance.  At $z\sim 100$, the Jeans scale is already mildly 
nonlinear, with $\delta M/M \sim 0.5$, causing 
the universe to already be substantially reionized ($f_{\II}>10\%$) 
at $z\sim 80$. But because of the early formation of non-linear structures, 
the clumping factor is higher, and hence the recombination rate 
is greater.  Full reionization is thus delayed until $z = 10\sim 20$.
Detailed calculations for reheating and reionization have not been done
for these models, so comparisons with other work can not be made.

\subsection{Current and Future Constraints on the CMBR: 
The Compton $y$ and the Optical Depth to Recombination}

The $y$-parameter measures the CMBR spectral distortion 
away from a blackbody spectrum caused by Compton-Thomson 
scattering.  Free electrons scatter and so Doppler shift 
CMBR photons, causing them to undertake a random walk
in frequency.  In thermal equilibrium, of course, the
spectrum does not shift, since by detailed balance, all
the processes must cancel.  However, after thermal 
decoupling at $z \sim 100$, the kinetic temperature of the gas 
evolves away from the CMBR 
temperature, first cooling by adiabatic expansion, and
then  heating by virializing in collaping structures or 
by photoheating from star formation.  The parameter
$y$ is defined by:
\begin{equation}
\dot{y} = \frac{k_B (T_e - T)}{m_e c^2} \sigma_T n_e c,
\label{eq:ydist_diffeq}
\end{equation}
where $T_e$ is the electron temperature, $T$ is the CMBR
temperature, and the other symbols have their usual meanings.
When structure formation begins in earnest and the gas 
temperature rises, low-energy CMBR photons 
will be scattered to higher energies.  

The $y$ we calculate in our models using 
equation~(\ref{eq:ydist_diffeq})
are shown in Table~\ref{tab:results}.
Again, the T+CDM and ICDM models behave similarly,
with $y = 2\sim 11 \times 10^{-7}$, and the 
Gaussian and T+ACDM
models have $y = 1\sim 3\times 10^{-7}$.  Thus, they
differ by a factor of $\sim 3$.
The COBE satellite 
\citep{Fixsen96}
put a bound on the spectral distortion
\begin{equation}
y\lesssim 1.5\times 10^{-5}.
\label{eq:ydist_maxbound}
\end{equation}
At long-wavelengths,
(in the Rayleigh-Jeans limit) 
the effective temperature $T_{\rm eff} = T + \delta T$
is lowered by a factor proportional to $y$:
\begin{equation}
\frac{\delta T}{T} = -2y.
\end{equation}
Unfortunately, the prospect for future experimental bounds on $y$ which
are significantly better than equation~(\ref{eq:ydist_maxbound}) 
are not good (Wilkinson, D., 1998, private communication).

As a check, an upper bound 
on $y$ can be obtained by assuming
that all the energy released by star formation
before redshift $z$ 
is transfered to the CMBR.  If  a fraction $d\epsilon$ 
of the baryon rest mass which has been
transfered, then the contribution the $y$-parameter, $dy$, is
governed by
\begin{equation}
d\epsilon\,\rho_b = 4 dy\, \rho_\gamma,
\end{equation}
or
\begin{eqnarray}
dy & =  & d\epsilon \frac{\rho_b}{4\rho_\gamma}\\
 & \approx & d\epsilon\frac{\Omega_b h^2}{10^{-4} (1+z)}
\label{eq:ydist_estimate}
\end{eqnarray}
where $\rho_\gamma$ is the photon energy (mass) density.  

Compton cooling is only efficient (i.e., faster
than the expansion rate) when $z \gtrsim 5$.
For the Gaussian models, the mass fraction in stars/quasars at this epoch 
is $f_\ast \sim 0.03$, and using the UV efficiency 
$\epsuv \approx 6\times 10^{-5}$ gives a bound
$\epsilon < 2 \times 10^{-6}$,
which implies that
$y < 4 \times 10^{-5}$.  This assumes that all the energy is
transferred at $z\sim 5$.  Integrating equation~(\ref{eq:ydist_estimate})
gives a value of $y < 0.6- 1.3 \times 10^{-5}$, all of them below the
COBE limit.  
The numbers for the T+CDM and ICDM models 
are about $y < 0.3-0.7 \times 10^{-5}$.  
The remaining factor of $\sim 10^{1.6-2.6}$ comes from 
several factors.  One is the fact that
the ionizing photons leave in the electrons on 
average $\sim 1-2$~eV 
per ionization in heat which can be transferred to the CMBR, 
resulting in a $\sim 10\%$ efficiency factor.
Furthermore,  there is competition from cooling processes other than 
Compton cooling, such as $\htwo$ and atomic line cooling.  
Thus the CMBR only absorbs a small fraction
of the total UV energy produced.
It appears, then, that the $y$-parameter is not a 
good distinguishing characteristic between these models.

A potentially stronger constraint on reionization
comes from the CMBR aniso\-tropy spectrum.  
The optical depth to recombination 
is given by
\begin{equation}
\tau_{\rm rec}  = \int n_e c \sigma_T dt
\end{equation}
On small angular scales, reionization results in the damping of
the temperature anisotropy spectrum by a factor $e^{-\tau_{\rm rec}}$.
Because we stop our calculation at $z=3$, we can only give a lower bound 
on the recombination optical depth.

\citet{zald98}
notes that while it might be difficult to 
extract the difference between a small degree of damping 
due to reionization and changes in other cosmological parameters,
even a modest optical depth due to reionization can enhance 
the polarization anisotropy spectrum on medium angular 
scales by an \emph{order of magnitude}.  Reionization causes
a polarization anisotropy because the polarization source 
is dominated by the temperature quadrupole, 
which grows after recombination via free-streaming 
of the temperature monopole.  
For example, 
\citet{zald98}
shows that an optical depth to recombination $\sim 0.5$
can produce a polarization anisotropy at $\ell=10$ on the order of
a micro-Kelvin ($\mu {\rm K}$), while the same model with 
$\tau_{\rm rec} = 0$  
gives a polarization anisotropy of of $\sim 10^{-1.5}\mu {\rm K}$. 

In the models we consider, those with the adiabatic CDM power
spectrum give a small optical depth of $\tau_{\rm rec} \lesssim 0.1$,
while the other models have $\tau_{\rm rec} \gtrsim 0.3$.  According
to Zaldarriaga, a negative detection on the $1^{\circ}$ scale
with sensitivity $\sim 1\mu{\rm K}$ would be able to rule out 
these moderate optical depth models.  The upcoming
MAP and PLANCK satellites may be able to distinguish between the 
low and high reionziation optical depths.  
As long as polarization foregrounds do not limit the measurements, 
the MAP and PLANCK satellites should be able to detect 
reionization optical depths of $\sim 0.02$ and $0.004$, respectively
(Spergel, D., private communication).

\subsection[Constraints on the Intergalactic Medium]
{Constraints on the Intergalactic Medium: 
The Gunn-Peterson Optical Depth and the Baryon Fraction}

Gunn-Peterson optical depth as a function
of scalefactor $a$ is given by \citep{PPC}
\begin{equation}
\tau_{\rm GP} = \frac{3\Lambda_{21} \lambda_{21}^2 \nhn(a) c}
	{4 H_0 a \sqrt{\Omega_0 a^{-3} + \Omega_\Lambda +
		(1 - \Omega_0 - \Omega_\Lambda) a^{-2}}},
\end{equation}
where $\Lambda_{21} = 6.25 \times 10^8\, \mbox{s}^{-1}$ 
is the Ly$\alpha$ decay rate and $\lambda_{21} = 1216$~\AA\ is the 
wavelength.  The Gunn-Peterson optical depth at $z=3$ is shown in
Table~\ref{tab:results}.  The results for all models 
are near the upper limits 
of $0.1 \sim 0.35$ at $z=4$ established by 
observations \citep{JO91,Giallongoetal94}.
Because we
have neglected the effect of large-scale 
shock heating (which would increase 
the temperature) as well as metals 
and other coolants (which would increase the star formation rate), 
it is plausible that our final neutral fractions are overestimated,
leading to an overestimate of $\tau_{\rm GP}$.

A stronger discriminant between models may be in their 
predicted star formation histories.  Figure \ref{fig:fstar}
shows the baryon fraction in
stars/quasars predicted by the various models we have considered.
Clearly, the early reheating models require that a significantly
greater fraction of the baryons be tied up in stars/quasars 
or their remnants.
The observational limit that we consider is from the 
analysis by  \citet{Rauchetal97}.  Their work
indicates that Ly$\alpha$ clouds account for the majority 
of the baryons in the universe.  
The strong limit on the Ly$\alpha$ clouds of 
$\Omega_{\rm cl} h^2 > 0.017$ combined with the most
generous (low deuterium abundance) big bang 
nucleosynthesis (BBN) constraint value of $\Omega_b h^2 = 0.024$,
implies that a fraction less than $0.017/0.024 = 0.29$ 
of the baryons in the universe can have been converted into stars/quasars
by a redshift $z = 3$.  
The models with the adiabatic power 
spectrum are well within this constraint, with stellar 
fractions $f_\ast < 0.06$.  
The T+ACDM models gives lower values
than the Gaussian models because of the lower value of
cluster normalization  required.
The ICDM model is within the limit
with $f_\ast = 0.2$; however, if the BBN value of $\Omega_b h^2$
falls below $\sim 0.02$, or if the limit is increased by more
than $\sim 15\%$, then the ICDM model will also encounter
difficulties.  The pure texture T+CDM 
gives stellar fractions which are at or above the limit
set by \citet{Rauchetal97}.

While we have used $\Omega_b h^2 = 0.0125$ in these models, 
we expect the \emph{fraction} 
of baryons converted to stars/quasars should be similar
or higher for models with higher $\Omega_b$.  The stellar
fraction depends on the fraction of baryons collapsing 
and on the fraction of collapsed mass which is cooling.
Increasing the baryon density will increase the cooling efficiency,
thus decreasing the Jeans mass and increasing the 
cooling fraction.  Both of these effects
will lead to an increased fraction of baryons 
undergoing star formation and thus an increased 
stellar fraction.  Of course, the mass-to-UV efficiency will 
need to decrease so as to produce the same $J_{21}$ at $z=4$.  

\section
{Discussion: Physical Bias and the Formation of Galactic Spheroids}
\label{sec:discussion}

In this section, we explore the relationship between 
luminous matter and the 
thermal and ionization history of the universe.
We begin by addressing the issue of physical bias 
at the Jeans mass limit 
using a generalized form of statistical
bias based on the peak-background split.  
We then consider the cooling efficiency of objects at 
or above the Jeans mass, and speculate as to the 
origin of galactic spheroids.

\subsection{Generalized Statistical Bias}

The peak-background split was first suggested by 
\citet{Kaiser84}
and recently revisited in the case of non-Gaussian models by
\citet{RGS98}.
In this model,
the density contrast field $\delta({\mathbf x})$ is split into
a short-wavelength ``signal'' $\delta_s$ and a 
long-wavelength ``background'' $\delta_b$.  The split is made in
such a manner that $\delta_s$ and $\delta_b$ are 
largely uncorrelated, $\langle \delta_s \delta_b\rangle \approx 0$.
The wavelengths of $\delta_s$ corresponds to the 
scales of the objects whose clustering properties are of interest,
and the wavelengths of $\delta_b$ corresponds to the scales 
at which the clustering properties (and bias) are to be measured. 
In this prescription, the local threshold for collapse $\delta_c$ is
modified locally to $\delta_c-\delta_b({\mathbf x})$, and
the (observed) Eulerian density is modified from the Lagrangian one
by a factor $1+\delta_b({\mathbf x})$.  
Let the average comoving number density of objects of 
scale $R$ to $R+dR$ be given by $\bar{N}_R(\delta_c)dR$, where
we have explicitly shown the dependence on the critical 
threshold of collapse $\delta_c$.  
Then the local comoving number density per unit scale $dR$ 
is given by
\begin{eqnarray}
N_R({\mathbf x}) & = & 
	\left[1 + \delta_b({\mathbf x})\right]\cdot
	\bar{N}_R\left[\delta_c-\delta_b({\mathbf x})\right]
	\\
& = & \left[1 + \delta_b({\mathbf x})\right]\cdot 
	\bar{N}_R\left(\delta_c\right) - 
		\delta_b({\mathbf x})
		\frac{\partial\bar{N}_R\left(\delta_c\right)}
			{\partial\delta_c}
		+ O\left[\delta^2_b({\mathbf x})\right],
\end{eqnarray}
where in the second line, we have used a Taylor expansion to
first order.  The long-wavelength ``bias at birth'' $b_\ast(R)$ 
can be defined by the ratio of the correlation 
functions at the time when the objects of scale $R$ are identified,
\begin{equation}
b_\ast^2(R) \equiv \frac{\xi_R({\mathbf x})}{\xi({\mathbf x})},
\end{equation}
where we have defined 
\begin{equation}
1 + \xi_R({\mathbf x}) \equiv 
	\frac{\langle N_R({\mathbf x})N_R({\mathbf 0})\rangle}
	{\bar{N}^2_R},
\end{equation}
and $\xi$ is the usual mass correlation function.  The bias at birth
is thus given by
\begin{equation}
b_\ast(R) = 1 - \frac{1}{\bar{N}_R}\frac{\partial \bar{N}}{\partial \delta_c}.
\end{equation}
The value of $\bar{N}_R$ is given from Press-Schechter 
by equations~(\ref{eq:ps_nrdef})--(\ref{eq:ps_alphadef}).  Differentiating
with respect to the critical threshold $\delta_c$ and rearranging
gives the following general form for $b_\ast(R)$:
\begin{equation}
b_\ast(R) = 1 + \frac{1}{\delta_c} \left[ 
	\left(\frac{-P'}{P}\right)\nu(R,t) - 1\right],
\label{eq:birthbias_generaldef}
\end{equation}
where $P' \equiv \frac{dP}{d\nu}$.  Examples of
the ratio $-P'/P$ are given in 
equations~(\ref{eq:pprime_p_gaus})--(\ref{eq:pprime_p_gaus}).
In the cases we consider, the bias at birth will be given by
\begin{equation}
b_\ast(R) = 1 +\frac{1}{\delta_c} \times
	\left\{\begin{array}{ll} \nu^2 - 1, & \qquad\mbox{Gaussian PDF}\\
	1.45\nu - 1, & \qquad\mbox{Texture PDF}\\
	0.67\nu - 1, & \qquad\mbox{Non-Gaussian ICDM PDF}
	\end{array}\right. .
\end{equation}
The Gaussian case has been derived previously by 
\citet{MW96} 
and others.  The non-Gaussian cases are similar to the derivation
for $\nu \gg 1$ by \citet{Kaiser84}.
It is important to note that for small $\nu$, an exponential
PDF such as that from textures or from the ICDM model gives a (slightly) 
larger bias than a Gaussian PDF; but for extremely rare
events with $\nu \gg 1$, a Gaussian PDF leads to a substantially
larger bias than exponential PDFs.  This was previously noted
by \citet{gooding92}. 

A similar calculation yields the bias of objects of 
scale $\geq R$ to be
\begin{equation}
b_\ast(\geq R) = \frac{\int_R^\infty dR\, \bar{N}_R\, b_\ast(R)}
	{\int_R^\infty dR\, \bar{N}_R} 
\end{equation}
As noted by \citet{MW96},
the total bias is simply the average of the individual biases, 
weighted by the mean number density of objects at each scale.
An equivalent form for $b_\ast(\geq R)$ 
was derived by \citet{RGS98}.

Below, we will primarily consider this bias at birth in the 
cosmological models we have analyzed.  This is equivalent to
considering models which are dominated by constant merging at 
the epochs we consider, so that objects do not
``survive'' long enough for their clustering bias to dynamically 
decrease.

\subsection{Physical Bias from Statistical Bias: 
Objects at the Jeans Scale}

As far as we know, 
a necessary condition for matter to be luminous is 
that the baryons collapse to high density.  Thus the first
criterion we use to explore physical bias is that objects
have a mass greater or equal to the Jeans mass.  

The Jeans birth bias $b_{\ast,J}$ is dependent on the PDF and the
degree on nonlinearity at the Jeans scale, $\nu_J$.  In 
Figure~\ref{fig:nuplot}, 
we show the value of $\nu_J$ in the ionized and neutral phases,
as well as the value of $\nu_J$ averaged between the two phases.
In most cases, the results are similar to those found by OG96:
during reheating, the average Jeans scale tracks the nonlinear 
scale defined by a constant value of $\nu_J$.  The early evolution of
$\nu_J$ is determined primarily by the power spectrum, and is
only weakly dependent on the PDF\null.  This is, of course, related
to the fact that the Jeans scale depends primarily on the
thermal evolution of the universe, which we found in the previous
section to be determined mainly by the power spectrum on 
small scales.  At epochs $z \lesssim 30$ all the models look 
quite similar.   This may be because of 
feedback between star formation and the 
temperature of the universe once reheating
has begun, keeping the Jeans scale at a constant nonlinear scale of 
$\nu = 2\sim 3.5$.  

The resulting bias is shown in 
Figure~\ref{fig:bplot}.
At high redshift, well before reionization, the Gaussian models 
exhibit the strongest bias, with $b_\ast > 10$ at 
redshifts $z> 30\sim 40$.  After a pause during reheating, 
during which $b_\ast$ remains in the range of $5\sim 7$, 
the bias drops quickly after full reionization to 
a level of $b_\ast = 1.5\sim 2$ at $z=3$. 
The T+ACDM model 
has an intermediate bias, falling to $b_\ast \sim 10$ at 
redshift $z \sim 80$, pausing during reheating at $b_\ast \sim 3$, 
and then falling to $b_\ast = 1.5\sim 2$ at $z=3$.  
The pure texture T+CDM and ICDM models
have a low bias which always remains $b_\ast \lesssim 3$, with the 
T+CDM bias declining slowly to from 2.5 to 1.5 after
reionization and the ICDM model bias always remaining $\sim 1$.

In contrast to many of the observational constraints considered
in the previous section, the evolution of $b_{\ast,J}$ 
depends on both the PDF and the fluctuation spectrum. 
While the value of $\nu_J$ depends primarily on the power spectrum,
the relationship between $\nu_J$ and $b_{\ast,J}$ is quite
different for different PDFs.

\subsection[Cooling Efficiency and Galactic Spheroids]
{Cooling Efficiency and the Origin 
of Galactic Spheroids}

The Jeans mass is the smallest possible baryonic 
mass which can collapse, and the value of $\nu_J$ derived here gives 
a lower bound on the average bias of all luminous objects.
But a second necessary condition which baryons must satisfy 
in order to become luminous is that they can cool, so as to 
collapse beyond virial density.  

From the cooling grid calculation described in 
\S~\ref{sec:gascooling_virial} and the prescription for
the cooling fraction $\epscool$ equation~(\ref{eq:epscool_def}), 
we can determine the constraints 
from cooling efficiency on the formation of
luminous objects.  In particular, 
at a given redshift $z$ and a minimum cooling fraction 
$\epsmin$, we can determine the maximum scale 
$R_c(z,\epsmin)$ for which halos collapsing at $z$
have $\epscool(R_c) \geq \epsmin$.  This gives the scale at
which at least a fraction $\epsmin$ of the baryons in the halo 
can cool.  Given the power spectrum
this maximum scale can be translated into a value for
the degree of nonlinearity $\nu_c(z,\epsmin)$.
To summarize, we have the correspondences:
\begin{eqnarray}
R_c(z,\epsmin) & \equiv & \mbox{max}[R],\ \mbox{such that}\  
	\epscool(R,z) \geq \epsmin\\
\nu_c(z,\epsmin) & \equiv & \frac{\delta_c}{\sigma[R_c(z,\epsmin)]}.
\end{eqnarray}
In Figures~\ref{fig:nu_j_cool_comp_flat}--\ref{fig:nu_j_cool_comp_open_icdm}
we show for $z\leq 30$ 
the values of $\nu_J$ derived in the previous subsection
along with the values of $\nu_c(z,\epsmin)$ for minimum
cooling fractions.  Each panel shows $\nu_J$ for the 
same value of $\Omega_0$, with 
different symbols for the different power spectra and PDF.
The lines show contours of $\epsmin = 10^{0,-0.5,-1}$, 
where all, $\sim 1/3$, and 1/10 of the baryons can cool.

What is particularly striking about the results of this 
calculation is that for $z \lesssim 10$, 
the ratio of the dynamical time to the cooling time for
Jeans mass objects is on the order of unity.  
Thus objects at or above the Jeans mass limit 
are \emph{barely} able to cool, or can only cool at their centers.  
The implications for galaxy formation are quite profound:
because the average cooling times and dynamical times are comparable, 
Jeans mass or larger gas clouds at redshifts $z \lesssim 10$ will on the average 
only have a marginal star formation efficiency.  The 
$\rho^{1/2} \sim r^{-1}$ dependence of the ratio of the dynamical
time to the cooling time means that the \emph{centers} of these
gas clouds will have significantly more efficient star formation.

The next question is whether the hierarchical nature of 
halo formation leads to gas clouds whose luminosity is 
dominated by orbiting stellar condensations or by a single central
stellar condensation.  In the first case, the gas cooling
efficiency of the more massive halos is small enough so that 
the cooling mass $M_c = M_b \epscool(z,M_b)$ begins to 
decrease when $M_b$ is greater than some critical value.  
In the second case, the cooling mass $M_c$ must be 
monotonically increasing with $M_b$.  

We find the second case to be more plausible 
because the dynamical to cooling ratio is a relatively weak 
function of mass.  
This fact is straight-forward to understand.  At redshifts $20 \geq z\geq 3$, 
Compton cooling is the dominant coolant at scales of
$\sim 1\hmpc$.  Because the Compton cooling coefficient $\propto (1+z)^4$, 
bremsstrahlung cooling does not become dominant until recent epochs. 
Thus at high temperatures, the cooling function 
$\Lambda_c \propto T$, and the cooling time $t_c \propto T/ \Lambda_c$ 
approaches a constant.  Since the dynamical time is determined by
the redshift, the ratio of the dynamical time to the cooling time
at high temperature is constant for all masses in a given epoch.  The
cooling \emph{fraction} thus approaches a 
constant, so $M_c \propto M$ and monotonically increases with 
the total mass $M$.  Even with brehmsstrahlung cooling, which rises 
as $\Lambda_c \propto T^{1/2}$, the cooling time would only decrease as 
$t_c \propto T^{-1/2}$.  The virial temperature is
related to the comoving scale by $T \propto R^2$, 
the cooling fraction will fall as $R^{-1}$.  Since the mass $M$
increases as $M\propto R^3$, the total cooling mass will still
increase as $M_c \propto R^2 \propto M^{2/3}$.

Therefore, the star forming regions of 
halos formed in the redshift range $10 \lesssim z \lesssim 3$ 
should always be dominated by their central star forming core
rather than orbiting stellar condensations leftover from mergers. 
Thus a picture of galactic spheroids forming at the centers of 
much more massive gas clouds and dark halos is in place. 

\section{Summary and Conclusions}
\label{sec:summary}

We have developed a semi-analytic model for the reheating and
reionization of the universe.  Our primary assumption is that UV
photons from early generations of star formation provide the
energy to heat and ionize the primeval gas.  We have 
explored the thermal and ionization evolution of the universe 
in a variety of cosmological models.  

Interestingly, we have found that 
many of the results found in a set of specific hydrodynamic simulations 
of the $\Lambda$-CDM model (GO97) appear to be generic:
(1) reheating precedes reionization, and clumping is important 
for understanding both processes; (2) 
full reionization is a sudden process of percolation and 
occurs near redshift $z = 12 \pm 5$; (3) 
during reheating, the Jeans mass closely tracks the nonlinear
mass scale; and 
(4) hierarchical clustering models are broadly consistent 
with the measured Gunn-Peterson opacity of the IGM.
However, models do have significant differences between which 
observations in the near future might be able to distinguish.

Our major conclusions are as follows:
\begin{enumerate}
\item
The most important determinant
of the thermal history of the universe for $z \leq 100$ 
is the power spectrum at the Jeans scale.  The shape of the
PDF is only of \emph{secondary} importance to the overall
thermal history.
For both Gaussian and texture PDFs, models with 
adiabatic CDM-type power
spectra, which have a logarithmically diverging fluctuation
spectrum $\sigma(R)$ at small scales, lead to 
reheating beginning at $z\sim 30$.  These we refer to
as ``late reheating models.''
For models with a power
law divergence at small scales, such as textures or ICDM, 
reheating has already begun at $z\sim 100$.
These we refer to 
as ``early reheating models.''

\item
In all models, full reionization occurrs as $z \sim 10$.
In models with late reheating, reionization proceeds after
reheating has plateaued.  In the early reheating models,
the increased clumping of gas caused by earlier 
structure formation delays full reionization until $z \sim 10$.  

\item
Two important observational constraints which may be
able to distinguish between different  models are the CMBR and the 
composition of the IGM\null.  
With regards to the CMBR, the polarization anisotropy 
caused by Compton scattering off free electrons will
be the strongest test of when the epoch of reionization occurred.
The late and early reheating models give different Compton 
optical depths to recombination which may be testable with experiments 
with sensitivity of $\sim 1 \mu$K\null.  With regards to the IGM, 
the strongest constraint is the baryon fraction in stars/quasars.  The
early reheating models have $> 20\%$ of the baryons sequestered
in stars/quasars by $z=3$, while the observational evidence suggests that
this fraction is smaller.  

\item
We found a correlation between the thermal and ionization 
history of the universe and the formation of luminous matter.
In particular, we found that during reheating, the Jeans mass 
scale tracks the nonlinear mass scale in all models, with 
$\delta M_J/M_J = 0.4\sim 1$, depending on the particular model.
After full reionization occurs, the Jeans scale remains roughly 
constant, so $\delta M_J/M_J$ grows.  

\item 
Since $\delta M/M$ can be related to galaxy bias, 
we find that bias at the Jeans scale
remains roughly constant during reheating, and then drops after
reionization.  Gaussian models generically give much higher biases
than the texture and non-Gaussian ICDM models, and is a potential
discriminant between models.

\end{enumerate}

We also considered the impact of the thermal and ionization 
history of the universe on galaxy formation.  It turns out that
Jeans mass objects at $z \lesssim 10$ are, on average, 
barely able to cool in 
a Hubble time.  Therefore, galaxy-sized objects will only cool
efficiently in their centers.  Galactic spheroids, then, may 
form at the centers of much larger, slowly cooling gas clouds.  
A rough picture of galaxy formation has thus emerged.

\acknowledgements{WAC and JPO acknowledge David Spergel and
David Wilkinson for helpful comments, and Nick Gnedin for
providing results from his simulations.  
WAC acknowledges David Spergel, David Wilkinson,
and Peter Meyers, who were all on his dissertation 
defense committtee. 
}

\appendix

\section{Thermal History of Neutral and Ionized Phases}

The temperature of the neutral (cold) phase will be determined 
primarily by adiabatic expansion and Compton scattering
of electrons off the CMBR\null.  We take the electron fraction
of the neutral phase $x_I$ to be equal to the residual electron
fraction $x_0$ from \citet{PPC} 
of 
\begin{equation}
x_I = x_0 \equiv \frac{1.2\times 10^{-5}\Omega_0^{1/2}}{h\Omega_b}
\end{equation}
where $\Omega_0$ is the present day value of the total
mass density parameter, $h$ is the Hubble parameter,
and $\Omega_b$ is the present day baryon density parameter.
We define the gas internal energy and particle number density
\begin{eqnarray}
u & \equiv & \frac{3}{2} n k_b T\\
n & \equiv & \frac{n_b}{\mu} \equiv \frac{n_H}{\mu X},
\end{eqnarray}
where $k_b$ is Boltzmann's constant, $T$ is the temperature,
$n$ is the total particle density, $n_b$ is the baryon density,
$\mu^{-1}$ is the particles per baryon, and $X$ is the
hydrogen mass fraction.  We have assumed an adiabatic index
$\gamma_a = 5/3$.

For 
simplicity, we consider only cooling
from hydrogen and electrons, using 
the compilation of \citet{anninosetal98}.
In erg cm$^{-3}$ s$^{-1}$, these rates,
followed by their reference and formulae (if not too complicated) are:
\begin{tabbing}
\hskip15pt \=
     $7.5\times10^{-19}(1+(T/{10^5})^{1/2})e^{-118348/T}~n_e n_{H^0}$ \kill
{\bf Compton cooling (heating)}: \citet{PPC}\\
\hskip15pt \>  $\Gamma_{cc} n_e \equiv 5.65\times 10^{-36} a^{-4} 
	\left(T - \frac{2.73}{a}\right)~n_e$\\
{\bf Collisional excitation cooling}: \citet{black81}; \citet{Cen92}
\\
\hskip15pt \>  $\Gamma_{ec} n_e \equiv 7.50\times10^{-19}~(1+\sqrt{T_5})^{-1}~
                \exp{(-118348/T)}~n_e n_{H^0}$\\
{\bf Collisional ionization cooling}: \citet{sk87}; \citet{Cen92}
\\
\hskip15pt \>  $\Gamma_{ic} n_e \equiv 
	2.18\times 10^{-11}~k_{ci} n_e n_{H^0}$ \\
{\bf Recombination cooling (Case A)}: \citet{black81};
\citet{Spitzer78}
\\
\hskip15pt \> $\Gamma_{rc} n_e \equiv 8.70\times10^{-27}~T^{1/2}~T_3^{-0.2}~
                (1+T_6^{0.7})^{-1}~n_e n_{H^+}$ \\
{\bf Molecular hydrogen cooling}:  Formula from \citet{leppshull83}
\\
\hskip15pt \> $\Gamma_{H_2c} n_e$ 
\\
{\bf Bremsstrahlung cooling}: \citet{black81};
\citet{spitzerhart79}
\\
\hskip15pt \>  $\Gamma_{bc} n_e \equiv 1.43\times10^{-27}~T^{1/2}
                ~[1.1+0.34\exp{(-(5.5-\log_{10} T)^2/3)}]
                ~n_e n_{H^+}$
\end{tabbing}
Here $k_{ci}$ is the H$^0$~+~e$^-$ collisional ionization rate coefficient 
taken from the 
\citet{Janev87}
compilation, and $T_n$ is the temperature
in units of $10^n$ Kelvin.  

We define the total cooling rate to be
\begin{equation}
\Gamma_{{\rm tot},c} \equiv 
\Cvar\, (\Gamma_{ec}+\Gamma_{ic}+\Gamma_{rc}+
	\Gamma_{H_2c} + \Gamma_{bc}) + \Gamma_{cc},
\end{equation}
where the clumping factor $\Cvar$ is required for the particle-particle
cooling rates, but not the Compton cooling rate.
The equations for thermal evolution in the H~I and H~II regions are thus
\begin{eqnarray}
\dot{u_I} &  =& - 5\frac{\dot{a}}{a}u_I 
	-\Gamma_{{\rm tot},c} x_I n_H\\
\dot{u_{\II}} &  =& - 5\frac{\dot{a}}{a}u_{\II} + \Gamma_{ph} (1-\xII) n_H 
	-\Gamma_{{\rm tot},c} \xII n_H,
\end{eqnarray}
where the photoheating rate 
$\Gamma_{ph}$ is given by equation (\ref{eq:gamma_ph}), 
and we have assumed that the electron density $n_e$ is equal to
the ionized hydrogen density $n_{H^+}$.  
The equations for the temperature evolution are thus
\begin{eqnarray}
k_b\dot{T_I} & = & - 2\frac{\dot{a}}{a} k_b T_I - \frac{2X}{3\mu}
	\Gamma_{{\rm tot},c} x_I
\label{eq:temp_i_diffeq}
	\\
k_b\dot{T_{\II}} & = & - 2\frac{\dot{a}}{a} k_b T_{\II} + \frac{2X}{3\mu}\left[
	\Gamma_{ph} (1-\xII) - \Gamma_{{\rm tot},c} \xII\right],
\label{eq:temp_ii_diffeq}
\end{eqnarray}
where $X$ is the hydrogen mass fraction. 

These temperatures are used to determine the comoving Jeans length
in each region:
\begin{equation}
R_{J,[I,\II]} = \frac{\pi}{k_{J,[I,\II]}} = 
	\sqrt{\frac{ 5\pi k_b T_{[I,\II]}}{12 G\bar{\rho}\mu m_H a^2}},
\end{equation}
where $k_J$ is the comoving Jeans wavenumber and
$\bar{\rho}$ is the \emph{total}
mean density (including dark matter).
The Jeans mass in baryons is thus
\begin{equation}
M_{J,[I,\II]} = \frac{4\pi}{3}\bar{\rho}_{0,b} R^3_{J,[I,\II]}
\end{equation}
where $\bar{\rho}_{0,b}$ is the current mean density in
baryons.

\section{Molecular Hydrogen Formation}

We follow the procedure described in \citet{anninosetal98}
to solve for the evolution of molecular hydrogen.  
The reactions involving molecular hydrogen are given in 
Table \ref{tab:h2reacts}.
The formation
of $\htwo$ is primarily through the intermediaries $\hm$ and $\htwop$,
which  \citet{anninosetal98}
note to be nearly in their 
(small) equilibrium abundances at all times.  
Neglecting the reaction $\htwop + \hm\rightarrow
\htwo + \hn,$ which is only a small order correction, 
the equilibrium abundance of $\hm$, $\xhm \equiv \nhm/n_H$, 
can be written independently of $\htwop$:
\begin{equation}
\xhm = \frac{k_7 \nhn x_e}
     {k_8 \nhn + k_{14}n_e + k_{15} \nhn + 
	(k_{16}+k_{17}) \nhp + k_{23}},
\label{eq:hm_abun_equil}
\end{equation}
where the variables $k_i$ are 
the rate coefficients with subscripts referring to
the reaction number in Table \ref{tab:h2reacts}.
Then given $\nhm$, the equilibrium abundance of 
$\htwop$, $\xhtwop \equiv \nhtwop/n_H$ can
be written with no additional assumptions as
\begin{equation}
\xhtwop = \frac{ k_9 \nhn \xhp      + k_{11}\nhtwo  \xhp
                 + k_{17}\nhm \xhp + k_{24}\xhtwo}
          {k_{10}\nhn + k_{18} n_e + k_{19}\nhm + k_{25} + k_{26}}.
\label{eq:htwop_abun_equil}
\end{equation}

The evolution equation for $\htwo$ is thus
\begin{equation}
\dot{x}_{\htwo} =  C - D\xhtwo,
\label{eq:htwo_diffeq}
\end{equation}
where the creation and destruction rates are
\begin{eqnarray}
      C & \equiv & k_8\nhm \xhn + k_{10}\nhtwop\xhn  + k_{19}\nhtwop\xhm
\label{eq:htwo_creation}
\\
      D & \equiv & k_{11}\nhp + k_{12}n_e + k_{13}\nhn +  
	k_{24}+k_{27}+k_{28}.
\label{eq:htwo_destruction}
\end{eqnarray}

\section{Gas Cooling in Virialized Halos}
\label{sec:gascooling_virial}

In Press-Schechter theory, 
the dynamical time for halos collapsing at $z_{\rm col}$ is
the same regardless of its mass, since the virialized 
density of the halo depends only on the collapse time.
Specifically, the dynamical time we use is:
\begin{eqnarray}
\tdyn(z_{\rm col}) & = & \sqrt{\frac{3\pi}{32 G\rho_{\rm vir}(z_{\rm col})}}
\label{eq:tdyn}
\\
\rho_{\rm vir}(z_{\rm col}) & = & 
	\Delta_{\rm vir}[\Omega(z_{\rm col})]\rho_{c}(z_{\rm col}),
\end{eqnarray}
where $\rho_c$ is the critical density, and the overdensity factor
$\Delta_{\rm vir}$ depends only on $\Omega$ at the time of 
collapse.

It is a much more involved process, however, to determine the
cooling time $\tcool$.  Although the mass density is the same
for all halos collapsing at a given epoch, different mass halos
have different virial temperatures, as well as different
temperature \emph{histories}.  Because the primordial chemistry
of the expanding universe not in equilibrium, we must follow
the thermal and ionization history of a halo in order to determine its
cooling time at virialization.  

Some previous semi-analytic models (e.g., 
Cole, Fisher, \& Weinberg 1994)
have simply assumed a 
collisional equilibrium cooling function 
$\Lambda_c(T) \equiv \Gamma_{{\rm tot},c}/n_H$ which
only depended on temperature.  
\citet{Tegetal97} considered the \emph{minimum} mass
which could cool within a Hubble time.  
Their results, as well as those of OG96 and GO97, 
showed that molecular hydrogen
plays a major role in the cooling of early halos
(as envisaged by Kashlinsky and Rees 1983 and Couchman and Rees 1986).
The formation
of $\htwo$, however, 
is both non-equilibrium and density dependent.  Thus, we
calculate a cooling rate for each halo, assuming a uniform 
spherical collapse of gas which ends up at the virial density.
Our cooling function depends on the temperature of the
halo and the collapse redshift.
We do make the simplifying assumption that the same cooling 
function $\Lambda_c(T,z)$ applies for the entire halo. 

To calculate $\Lambda_c(T,z)$, we extensively 
modified the non-equilibrium
``T0D'' ionization code of
\citet{Abel98} so that it 
modeled a spherical
collapse.  This code incorporates all of the physics of 
 \citet{anninosetal98}
and \citet{Abeletal98}. 
We assume a uniform 
sphere of gas and dark matter which collapses to one-half its
radius of maximum expansion and which heats the gas to the
virial temperature.  Before turnaround, we assume the standard
parametric form for spherical collapse, given by:
\begin{eqnarray}
r & = & A(1 - \cos\theta)\\
t & = & B(\theta - \sin\theta)\\
A^3 & = & G M B^2\\
t_{\rm ta} & = & \pi B\\
r_{\rm max} & = & 2A,
\end{eqnarray}
where $t_{\rm ta}$ is the time at turnaround,  
$r_{\rm max}$ is the maximum radius of expansion,
and $M$ is the total mass of the halo.
After turnaround, we adopt the following simple functional form for the 
radius as a function of time:
\begin{eqnarray}
r & = & A\left(2 - 
	\frac{\tau^2}{\tau^2 + 8\, {\rm sech} \frac{1}{2}\tau^2}\right)\\
\tau & \equiv & \frac{t - t_{\rm ta}}{B}
\end{eqnarray}
For the velocity dispersion $\sigma^2$ 
of the gas, we assume shock heating leads to a rise from its 
value at turnaround $\sigma^2_{\rm ta}$ to the virial velocity 
dispersion $\sigma^2_{\rm vir}$ via the equation
\begin{equation}
\sigma^2 = \sigma^2_{\rm ta} + (\sigma^2_{\rm vir}-\sigma^2_{\rm ta})\cdot
	\left(1 - {\rm sech}\frac{1}{2}\tau^2\right).
\end{equation}
Plots of these functions are shown in Figure \ref{fig:sphcolfunc}.

The calculated value of the ratio $\tdyn/\tcool$ for gas at the
virial density is given in 
Figures~\ref{fig:coolgrid_Omega_one} and 
\ref{fig:coolgrid_Omega_zerothreefive}, 
for the cases 
of $\Omega_0 = 1.0$ and $\Omega = 0.35$.  Also shown is
the comoving Jeans length for gas in the neutral phase.
The narrow ridge in which the cooling time is longer than the dynamical time 
extending from $R \approx 0.02,\ z \approx 25$ to $R\approx 0.01,\ 
z \approx 100$ for the $\Omega_0 =1$ case, and 
$R \approx 0.04,\ z \approx 20$ to $R\approx 0.02,\ 
z \approx 100$ for $\Omega_0 = 0.35$, 
corresponds to when $\htwo$ has been 
destroyed, but atomic line cooling is not yet efficient.
This ``pause'' in the cooling efficiency was first identified
by \citet{OG96}.
Note that $\htwo$ cooling remains efficient later for $\Omega_0 = 0.35$
than for $\Omega_0 = 1$ (the tip of the ``banana'' shaped region to the left
of the ridge extends to lower redshift ).

This calculation is essentially a simplified version of the
\citet{Tegetal97}
calculation, but we use the information
in the \emph{entire} ``cooling grid'' in our work.  This is
particularly important at later times, since the cooling efficiency
actually begins to decrease for large masses (this is why we have 
clusters of galaxies rather than gigantic $\sim 10^{15}\, M_{\odot}$
mega-galaxies).  These ``cooling grids''
can be precalculated for a given background cosmology, and
are assumed to be the same for any PDF.  

Even if the average cooling time of a halo is greater
than or equal to the dynamical time, it is possible that
the central, densest parts of a halo can still cool.  
To see why, consider a constant cooling function $\Lambda_c$.
Because the cooling time is inversely proportional to the 
density $\tcool \propto \rho^{-1}$, and the 
dynamical time $\tdyn \propto \rho^{-1/2}$, the
ratio of the dynamical to the cooling times is proportional to
$\rho^{1/2}$:
\begin{equation}
\varphi \equiv \frac{\tdyn}{\tcool} \propto \rho^{1/2}
\end{equation}
If we make the \emph{ansatz} that only regions cooling on a timescale
shorter than the dynamical timescale, so that 
$\varphi \geq 1$, can undergo star formation, then 
there will be a maximum radius, possibly equal to the virial
radius, outside of which star formation will not occur.
For a singular isothermal
sphere, the density is proportional to $r^{-2}$, and the
mass within a radius $r$ is proportional to $r$.
Therefore the fraction 
of the baryonic mass of the halo undergoing
star formation is simply proportional to $\varphi$:
\begin{equation}
\frac{M_b[r(\varphi = 1)]}{M_{b,\rm vir}}  =  
	\frac{r(\varphi=1)}{r_{\rm vir}} =  
	\sqrt{\frac{\rho_{\rm vir}}{3\rho(r)}} \propto \varphi.
\end{equation}
If we define $\varphi_0$ to be the value of $\varphi$
for gas at the virial density of the halo, then
since the mean density of the halo $\rho_{\rm vir}$ equals the 
local density when $r = r_{\rm vir}/\sqrt{3}$,
the fraction of a halo which is cooling faster than
the dynamical timescale 
$\epscool(M_b)$ will
be given by
\begin{equation}
\epscool(M_b) = \mbox{min}\left[\frac{\varphi_0}{\sqrt{3}},\ 1\right].
\label{eq:epscool_def}
\end{equation}
This is, of course, an upper limit, since halos will have cores
so that the density no longer scales as $r^{-2}$.  
To account for core radii, we set a $\epscool = 0$ if 
$\varphi_0/\sqrt{3} < 0.01$, 
corresponding to a core radius equal to about 1/30 of the
virial radius.

\section{Calculational Implementation}

\subsection{Scaled Equations}

In order to simplify our differential equations 
(\ref{eq:dnxdt_new}) and (\ref{eq:dedtraw_new}), 
we convert to 
dimensionless energy densities by dividing $\bar{E}$ and $\bar{S}$ by
$\epsilon_0 n_H$:
\begin{eqnarray}
\Evar & \equiv & \frac{\bar{E}}{\epsilon_0 n_H}
\\
\Svar & \equiv & \frac{\bar{S}}{\epsilon_0 n_H}.
\end{eqnarray}
We scale the luminosity $\ell(M_b,\ttilde,t)$ equation 
(\ref{eq:lumeq}) by $\epsilon_0$:
\begin{eqnarray}
\Lvar(M_b,\ttilde,t) & \equiv & \frac{\ell(M_b,\ttilde,t)}{\epsilon_0}
	\nonumber\\
	& = &
	\left[1 - f_\ast(\ttilde)\right] 
	\frac{M_b \epscool \epseff c^2}{\epsilon_0\ttdyn}
	\left(\frac{t-\ttilde}{\ttdyn}\right)
	\exp\left[-\left(\frac{t-\ttilde}{\ttdyn}\right)\right].
\end{eqnarray}
We also make the following definitions:
\begin{eqnarray}
n_0 & \equiv & n_H a^3\\
\zeta & \equiv & \Evar \lambda^{-1/3}\\
\beta & \equiv & n_H c \sigma_p\\
\xi & \equiv & n_H c \sigma_e\\
\Gamma & \equiv & \Cvar n_H \alpha_R\\
K_{ci} & \equiv & \Cvar n_H k_{ci} \\
H & \equiv & \frac{\dot{a}}{a}\\
\eta^{-1} & \equiv & \xII(1-\fII) = \xII e^{-\lambda}.
\end{eqnarray}
The expression for $\Svar$ in closed form from 
equation (\ref{eq:sbar_int_l}) is thus
\begin{equation}
\Svar   = n_0^{-1}
	\int_0^{\infty} dM_b \int_0^t d\ttilde\,
	\psi(M_b,\ttilde) \cdot
	N(M_b,\ttilde,t)\cdot \Lvar(M_b,\ttilde,t);
\label{eq:svar_closedform}
\end{equation}
the expression for $\zeta$, using equation 
(\ref{eq:ebar_lambda_rel}), is
\begin{equation}
\zeta = \frac{a}{c}\left(\frac{3}{4\pi \Svar}\right)^{1/3}
	n_0^{-4/3} \int_0^\infty dM_b \int_0^t  d\ttilde\,
	\psi(M_b,\ttilde)\cdot N(M_b,\ttilde,t)\cdot 
	\Lvar^{4/3}(M_b,\ttilde,t).
\label{eq:zeta_closedform}
\end{equation}
Note that although the integrands which determine 
$\Svar$ and $\zeta$ depend on $\fII$ through the
selection function $\psi$, this dependence is only for  
$\fII$ evaluated at times $\ttilde < t$
since $\Lvar(M_b,\ttilde=t,t) = 0$.  

These integrals can be simplified by making the definition
\begin{eqnarray}
\dot{\varrho}(\ttilde) & \equiv & \int dR\, \psi(R,\ttilde)
	\dot{N}_{\rm form}(R,\ttilde) \times 
	\nonumber \\
	& & \qquad \left[ 1 - f_\ast(\ttilde)\right]
	\epscool(R,\ttilde) \frac{4\pi R^3}{3}
	\left(\frac{m_H c^2 \epseff}{X \epsilon_0}\right),
\end{eqnarray}
which is the cumulative source function for star formation, 
with energy in Rydbergs.  The scaled source function $\Svar$
is given by smoothing $\dot{\varrho}$ over a dynamical time,
and accounting for the destruction of halos through mergers.
The smoothing function $\Kvar(\ttilde,t)$, from the 
definition of $\ell$, is given by
\begin{equation}
\Kvar(\ttilde,t) \equiv \frac{1}{\ttdyn} 
	\left(\frac{t-\ttilde}{\ttdyn}\right)
	\exp\left[-\left(\frac{t-\ttilde}{\ttdyn}\right)\right]
	\frac{D(\ttilde)}{D(t)},
\end{equation}
so the scaled source by
\begin{equation}
\Svar(t) = \int^t d\ttilde\,  \Kvar(\ttilde,t) \dot{\varrho}(\ttilde).
\end{equation}
Similarly, defining 
\begin{eqnarray}
\dot{\varrho}_{4/3}(\ttilde) & \equiv & \int dR\, \psi(R,\ttilde)
	\dot{N}_{\rm form}(R,\ttilde) \times 
	\nonumber \\
	& & \qquad \left\{ 
	\left[ 1 - f_\ast(\ttilde)\right]
	\epscool(R,\ttilde) \frac{4\pi R^3}{3}
	\left(\frac{m_H c^2 \epseff}{X \epsilon_0}\right)
	\right\}^{4/3},
\end{eqnarray}
we obtain from 
equation~(\ref{eq:zeta_closedform})
\begin{equation}
\zeta = \frac{a}{c}\left(\frac{3}{4\pi \Svar}\right)^{1/3}
\int^t d\ttilde\, \Kvar(\ttilde,t) \dot{\varrho}_{4/3}(\ttilde).
\end{equation}

In terms of the scaled source function $\Svar$, 
the differential equation for $f_\ast$, 
the fraction of baryons in stars/quasars, is:
\begin{equation}
\dot{f}_\ast = \epsstar \frac{\Svar(t)\,\epsilon_0 X}{\epseff m_H c^2},
\end{equation}
obtained by differentiating equation (\ref{eq:fstar_int_eq}).
The ionization fraction and UV energy evolution 
equations (\ref{eq:dnxdt_new}) and (\ref{eq:dedtraw_new})
thus become
\begin{eqnarray}
\dot{\xII} & = & \xII\left[(1 - \xII) K_{ci} -\xII \Gamma\right]+ 
	(1 - \xII)\frac{\beta\Evar}{\fII}
\label{eq:xdot_full}
\\
\dot{\Evar} & = & \Svar - \left[\gamma H  + (1-\xII)\xi\right]\Evar  
	- \eta^{-1} \dot{\lambda},
\label{eq:evardot_full}
\end{eqnarray}
having replaced $\dot{\fII}$ by $(1-\fII)\dot{\lambda}$.
Our constraint equation, derived from 
equation~(\ref{eq:ebar_lambda_rel}), is  
\begin{equation}
\lambda = \left(\frac{\Evar}{\zeta}\right)^3,
\label{eq:lambda_cons_full}
\end{equation}
and $\Svar$ and $\zeta$ are defined above.

\subsection{Finite Difference Equations}

The differential equations (\ref{eq:xdot_full}) and 
(\ref{eq:evardot_full}) are
non-linear and ``stiff,'' so forward differencing such as by 
Runge-Kutta is not stable.  
Two methods often used in such stiff systems are 
``implicit'' and ``semi-implicit'' differencing 
\citep{NR92}.
Implicit differencing evaluates the right-hand side derivatives
at the new location, but is only guaranteed to have a stable solution
in closed form for linear systems.  In our case, the nonlinearity of
the equations precludes using purely implicit differencing.
Semi-implicit differencing linearizes the right-hand side derivatives, 
inverting a matrix at each step, but is not guaranteed to be stable.
Furthermore, because of the complicated nature of the
right-hand side derivatives, the Jacobian matrix required
for these methods is extremely cumbersome to evaluate.
Below, we use a simple ``mostly backward'' differencing scheme.

At each timestep, we first evaluate $\Svar$ and $\zeta$ 
using equations (\ref{eq:svar_closedform}) and (\ref{eq:zeta_closedform}),
and update the temperature of each phase using
equations (\ref{eq:temp_i_diffeq}) and (\ref{eq:temp_ii_diffeq}).
At very high redshift, $\Svar$ and $\zeta$ will be zero;
we  leave $\xII$ at the post-recombination residual
ionization, and $\Evar$ and $\lambda$ as zero.
When $\Svar$ and $\zeta$ become nonzero, we 
solve for the new values of $\lambda$ and $\Evar$
simultaneously from equations (\ref{eq:evardot_full}) and 
(\ref{eq:lambda_cons_full}), using the previous timestep
value for $\xII$. 
We then solve the ionization 
equation (\ref{eq:xdot_full}) to update the value of
$\xII$.  Finally, we update the abundance of $\htwo$.
The finite differencing implementation of this procedure
follows below.

In differencing equation  (\ref{eq:evardot_full}),
we take the values of $(1-\xII)$ and $\eta$ at the current
timestep $t_i$.  The other
values are evaluated at the advanced time step $t_{i+1}$.  
The finite difference equation for $dt = t_{i+1}-t_i$ is thus
\begin{equation}
\Evar_{i+1} - \Evar_i = dt\, \Svar_{i+1}
	- dt [\gamma\, H_{i+1} + (1-\xII)_i\xi_{i+1}]\,\Evar_{i+1}
	- \eta_i^{-1}(\lambda_{i+1}-\lambda_i).
\end{equation}
Replacing $\lambda_{i+1}$ with $(\Evar_{i+1}/\zeta_{i+1})^3$, 
equation (\ref{eq:lambda_cons_full}), and collecting terms gives 
the equation
\begin{equation}
\left(\frac{\Evar_{i+1}}{\zeta_{i+1}}\right)^3\eta_i^{-1} 
	+ \Evar_{i+1} U - V = 0
\label{eq:evar_next_full}
\end{equation}
with
\begin{eqnarray}
U & \equiv & 1 + dt\,\gamma\,H_{i+1} + 
		dt\,(1-\xII)_i\xi_{i+1}
\label{eq:udef_full}
\\
V & \equiv & \lambda_i\eta_i^{-1} + \Evar_i + \Svar_{i+1} dt.
\label{eq:vdef_full}
\end{eqnarray}
While this cubic equation has an algebraic solution, it is
faster to solve for 
$\Evar_{i+1}$ iteratively using Newton's method.
Equation (\ref{eq:evar_next_full}) is of the form
\begin{equation}
f(y) \equiv Cy^3 + Uy - V = 0,
\end{equation}
where $y = \Evar_{i+1}$.
Newton's method iterates to a solution using 
\begin{equation}
y_{j+1}  =   y_j - \frac{f(y_j)}{f'(y_j)}
	 =  y_j - \frac{C y_j^3 + Uy_j - V}{3C y_j^2 + U}.
\end{equation}
We use the subscript $j$ so as to not confuse this iterative
root finding with our time step labels $i$.
The key to Newton's method is a good ``first guess'' for $y$.
Since $U,\, V,\, C\, >\, 0$, we know that the solution $\tilde{y}$ satisfies
$\tilde{y} < (V/C)^{1/3}$ and $\tilde{y} < V/U$.  So our first
guess $y_0 = \mbox{min}[(V/C)^{1/3},V/U]$.  In cases where
$U \ll C^{1/3} V^{2/3}$ or $U \gg C^{1/3} V^{2/3}$, the iteration will 
converge in only one or two steps.
Once a solution for $\Evar_{i+1}$ is found, we then use equation 
(\ref{eq:lambda_cons_full}) to find $\lambda_{i+1}$.

We use the updated values for $\Evar$ and $\lambda$ 
in the ionization equation (\ref{eq:xdot_full}).
We evaluate all values at the advanced time step.  
The finite difference equation is thus
\begin{eqnarray}
x_{\II,i+1} - x_{\II,i} & = &
	dt\, \biggl[x_{\II,i+1}(1 - x_{\II,i+1}) K_{ci,i+1}
  	- x^2_{\II,i+1} \Gamma_{i+1} \nonumber \\
	& & \quad\quad
	+ (1 - x_{\II,i+1})\beta_{i+1} 
	\frac{\Evar_{i+1}}{f_{\II,i+1}}\biggr].
\end{eqnarray}
Collecting terms gives
\begin{eqnarray}
& & x^2_{\II,i+1} dt(\Gamma_{i+1} + K_{ci,i+1}) + 
	x_{\II,i+1} 
	\left[1 + dt\left(\beta_{i+1} 
		\frac{\Evar_{i+1}}{f_{\II,i+1}}
		- K_{ci,i+1}\right)\right] = \qquad
\nonumber \\
& & \qquad\qquad\qquad\qquad\qquad x_{\II,i}  
 + dt\,\beta_{i+1} \frac{\Evar_{i+1}}{f_{\II,i+1}},
\end{eqnarray}
or equivalently
\begin{equation}
x^2_{\II,i+1} + A\, x_{\II,i+1} - B = 0
\end{equation}
with
\begin{eqnarray}
A & \equiv & 
	\frac{ 1 + dt\left(\beta_{i+1} 
		\Evar_{i+1}/f_{\II,i+1} - K_{ci,i+1}\right)}
	{dt\left(\Gamma_{i+1} + K_{ci,i+1}\right)}
\\
B & \equiv &
	\frac{x_i +  dt\,\beta_{i+1} 
		\Evar_{i+1}f_{\II,i+1}}
	{dt\left(\Gamma_{i+1} + K_{ci,i+1}\right)}.
\end{eqnarray}
This can be solved for $x_{\II,i+1}$ either algebraically or 
iteratively.

The new values of the electron fraction are used to
update the molecular hydrogen abundance.
Converting the differential equation for the evolution of $\htwo$ 
equation (\ref{eq:htwo_diffeq}) to a backward 
difference equation gives
\begin{equation}
      x_{\htwo,i+1}  =  \frac{x_{\htwo,i} + C_{i+1}\,dt}{1 + D_{i+1}\,dt}
\end{equation}
where the creation $C$ and destruction $D$ rates are given by
equations (\ref{eq:htwo_creation}) and (\ref{eq:htwo_destruction}. 
We use the equilibrium abundances of 
$\hm$ and $\htwop$, equations 
(\ref{eq:hm_abun_equil})--(\ref{eq:htwop_abun_equil}).

\clearpage

\figcaption[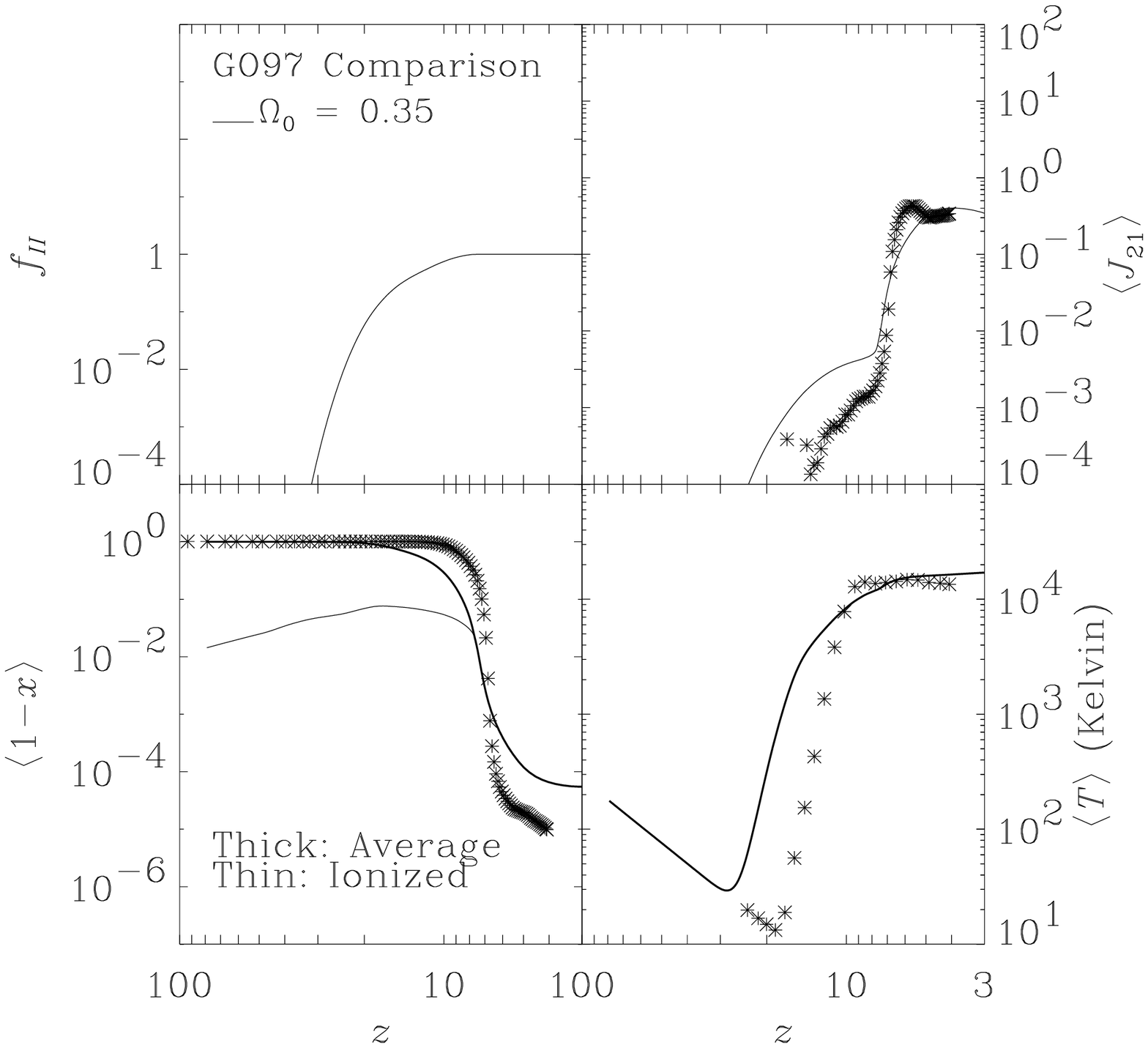]{
Results of reionization calculation with GO97 parameters:
$f_{II}$ is the filling factor; $\langle J_{21} \rangle$ is
the global average of the ionizing intensity $J_{21}$; 
$\langle 1-x\rangle$ is the neutral fraction (thick: global 
average, thin: ionized 
region); and $\langle T \rangle$ is the temperature.
The results from the GO97 
simulation are denoted by the stars.\label{fig:go_comp_fourplot}
}

\figcaption[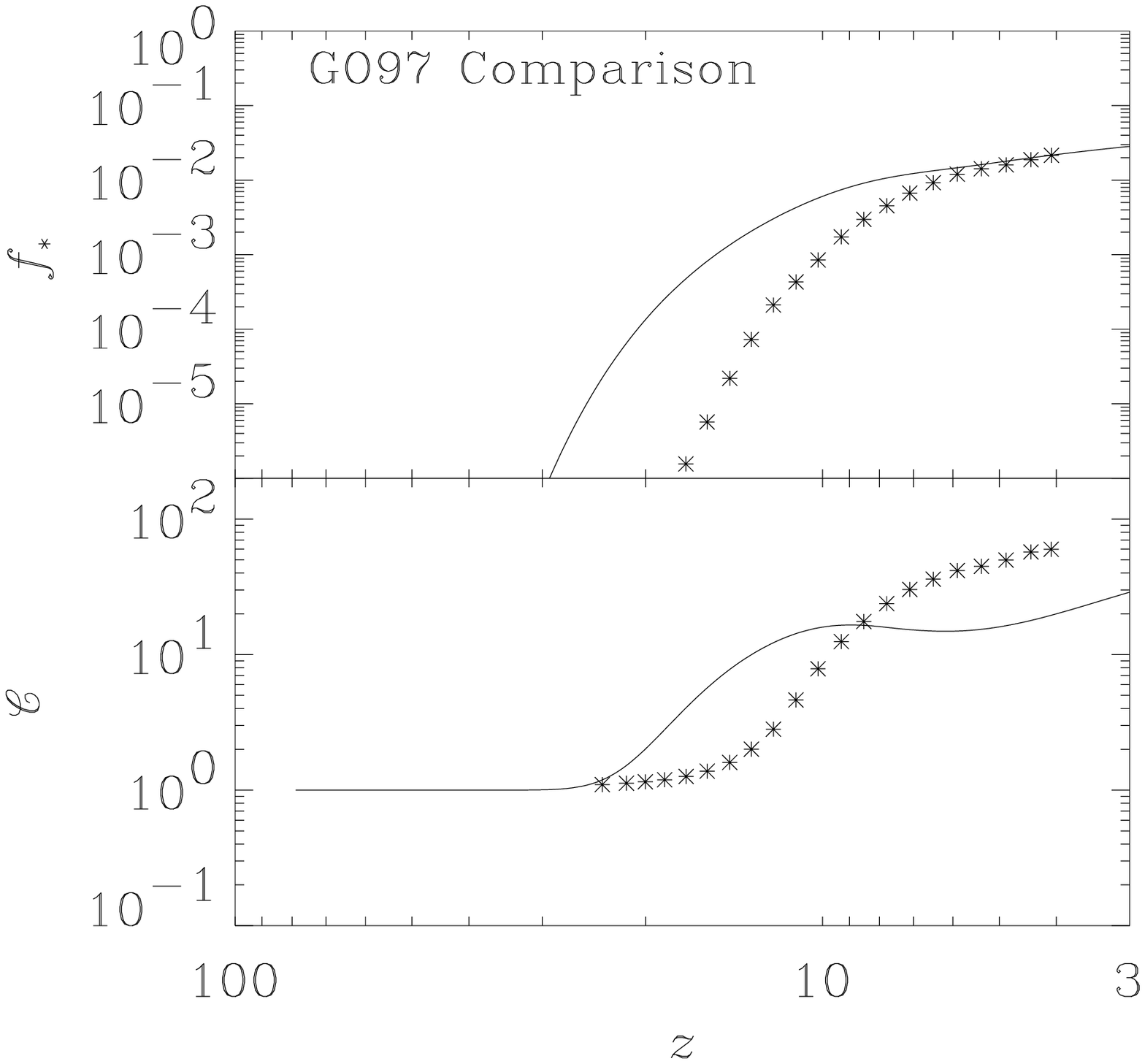]{
The fraction of baryons in stars/quasars $f_\ast$,
and the clumping factor $\Cvar$ for reionization 
calculation with GO97 Parameters.
The results from the GO97 
simulation are denoted by the stars.
\label{fig:go_comp_twoplot}
}

\figcaption[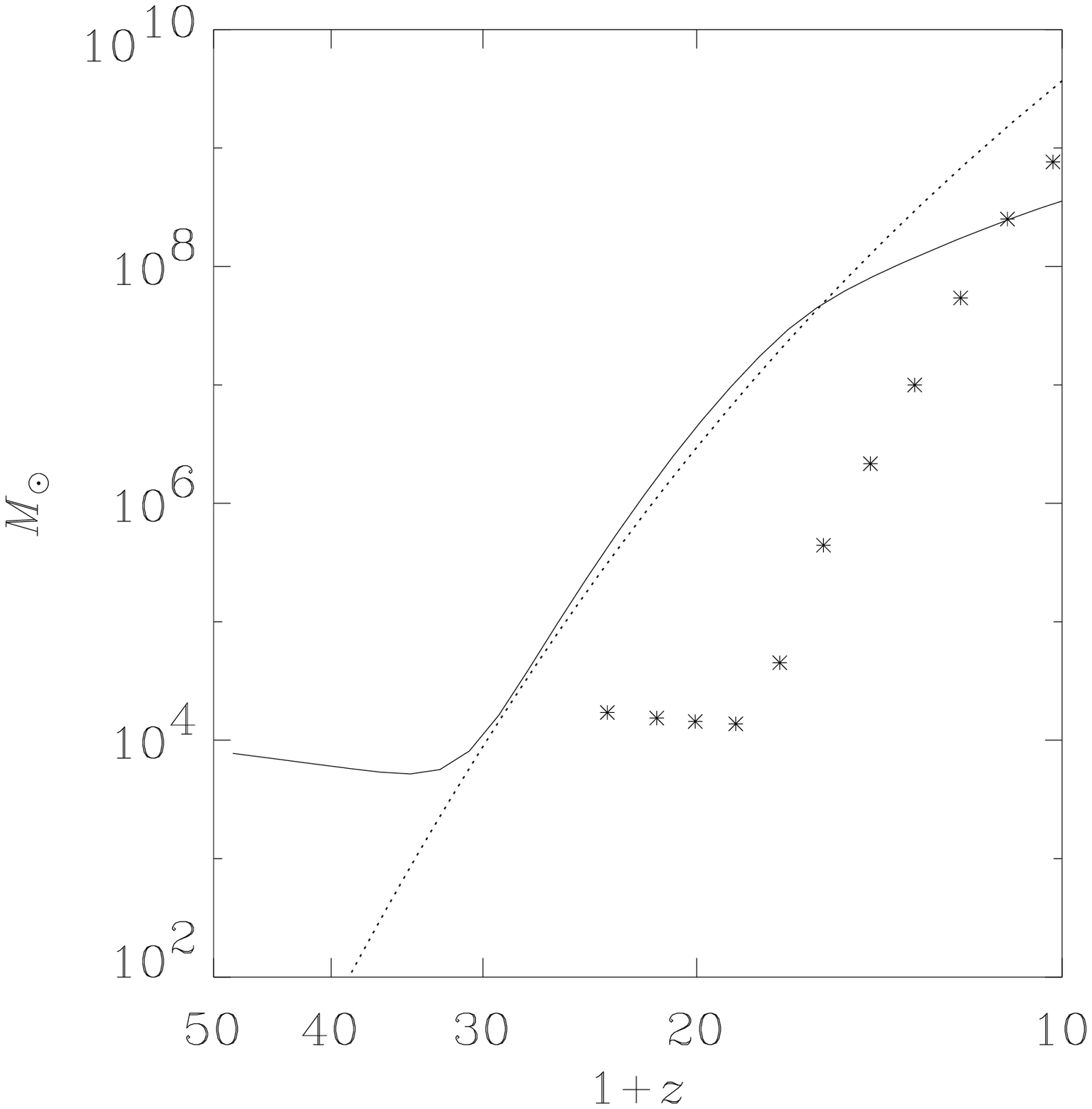]{
Jeans and nonlinear mass scales for GO97 comparison: solid line marks
the volume-averaged Jeans mass in baryons 
derived from our semi-analytic model; the dotted line
denotes the mass scale at which $\delta M/M \approx 0.4$ at each epoch.
The Jeans mass from the GO97 simulation, as described in 
\citet{OG96}, is denoted by the stars.
\label{fig:go_comp_mj_mnl}
}

\figcaption[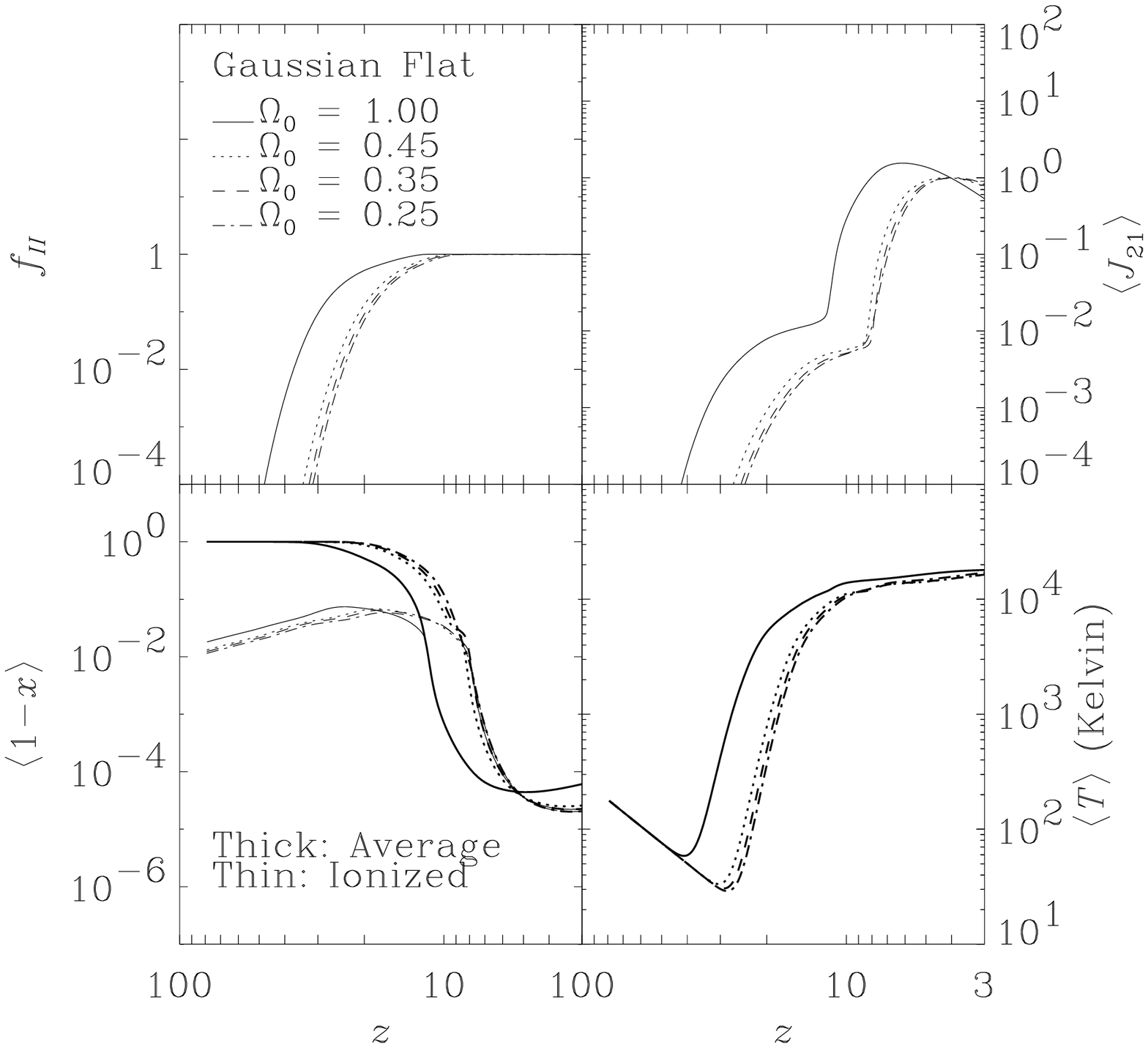]
{
Results of reionization calculation for flat Gaussian models G+ACDM:
$f_{II}$ is the filling factor; $\langle J_{21} \rangle$ is
the global average of the ionizing intensity $J_{21}$; 
$\langle 1-x\rangle$ is the neutral fraction (thick: global 
average, thin: ionized 
region); and $\langle T \rangle$ is the temperature.
\label{fig:gaus_flat_fourplot}
}

\figcaption[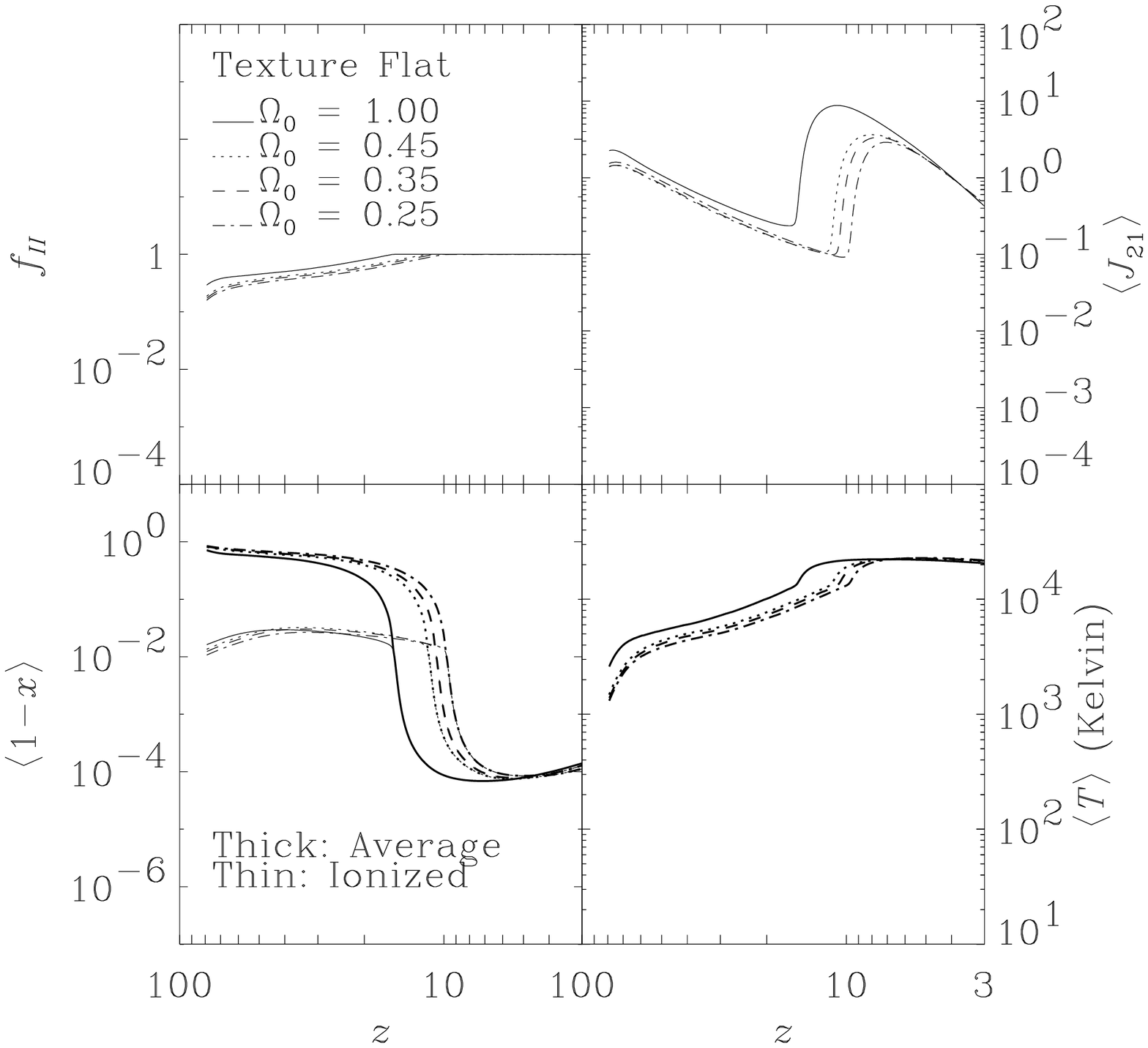]
{
Results of reionization calculation for flat texture models
T+CDM.
\label{fig:text_flat_fourplot}
}

\figcaption[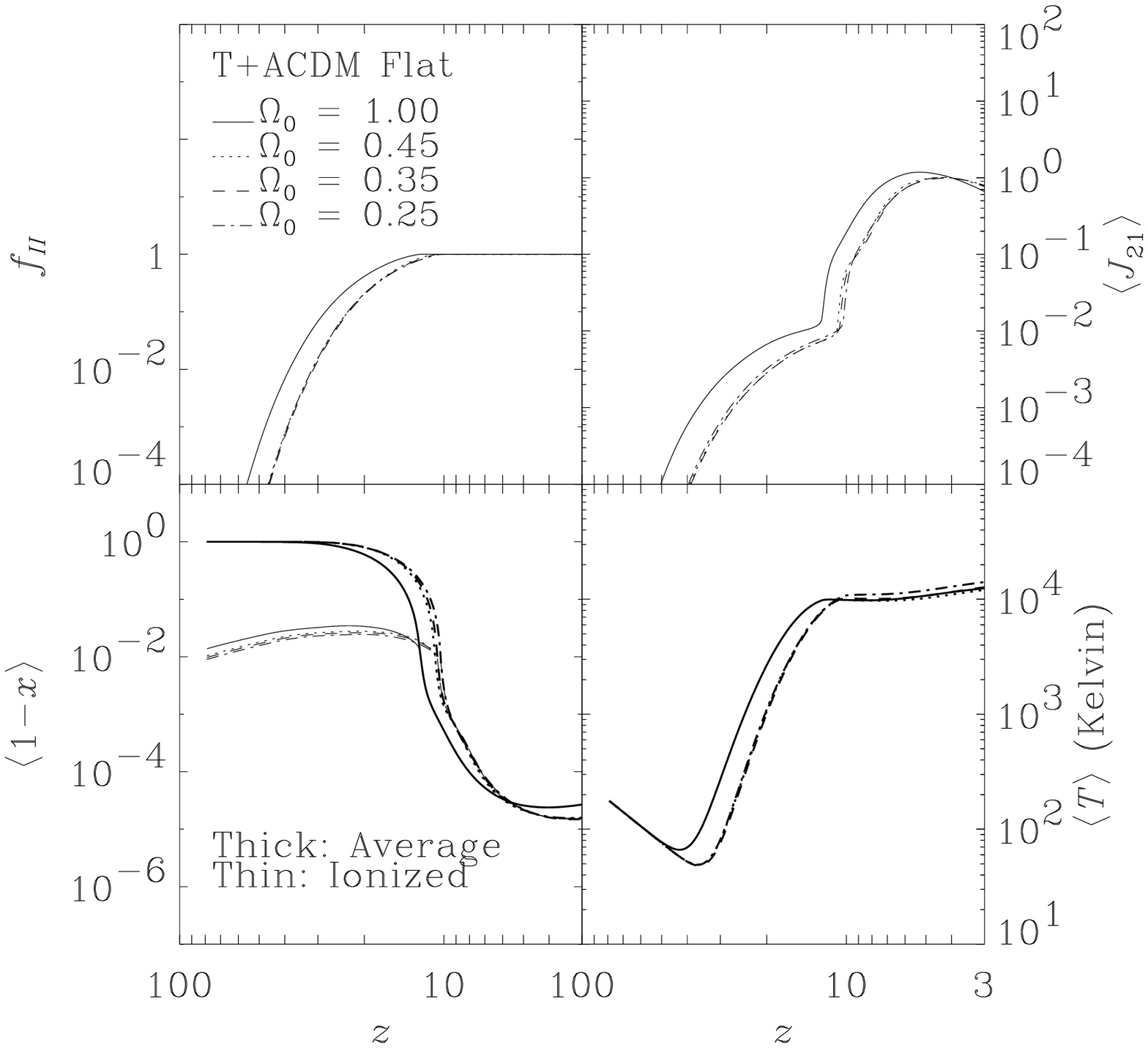]
{
Results of reionization calculation for flat texture models
with adiabatic CDM power spectrum T+ACDM.
\label{fig:text_flat_cdmpk_fourplot}
}

\figcaption[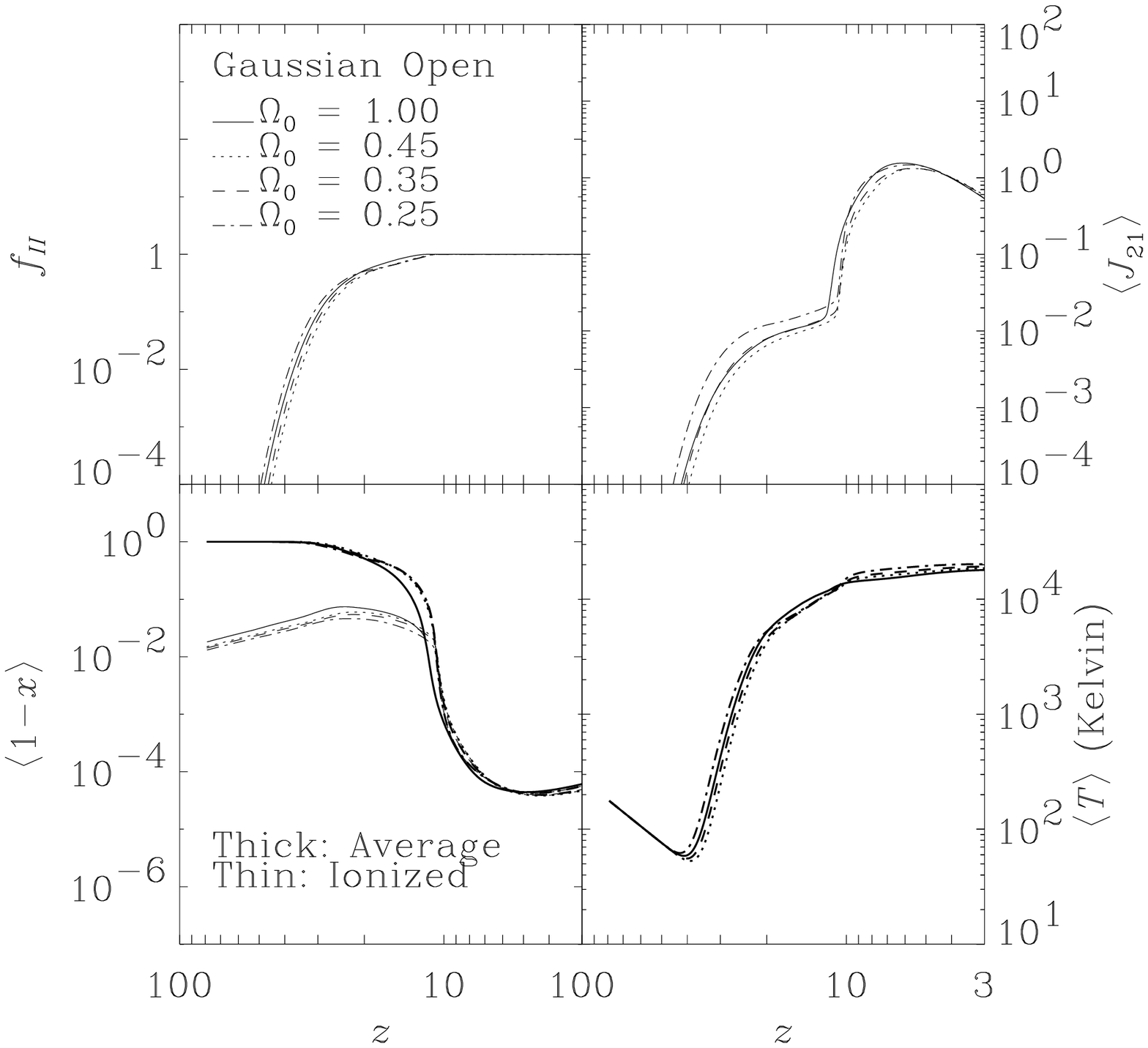]
{
Results of reionization calculation for open Gaussian models.
\label{fig:gaus_open_fourplot}
}

\clearpage

\figcaption[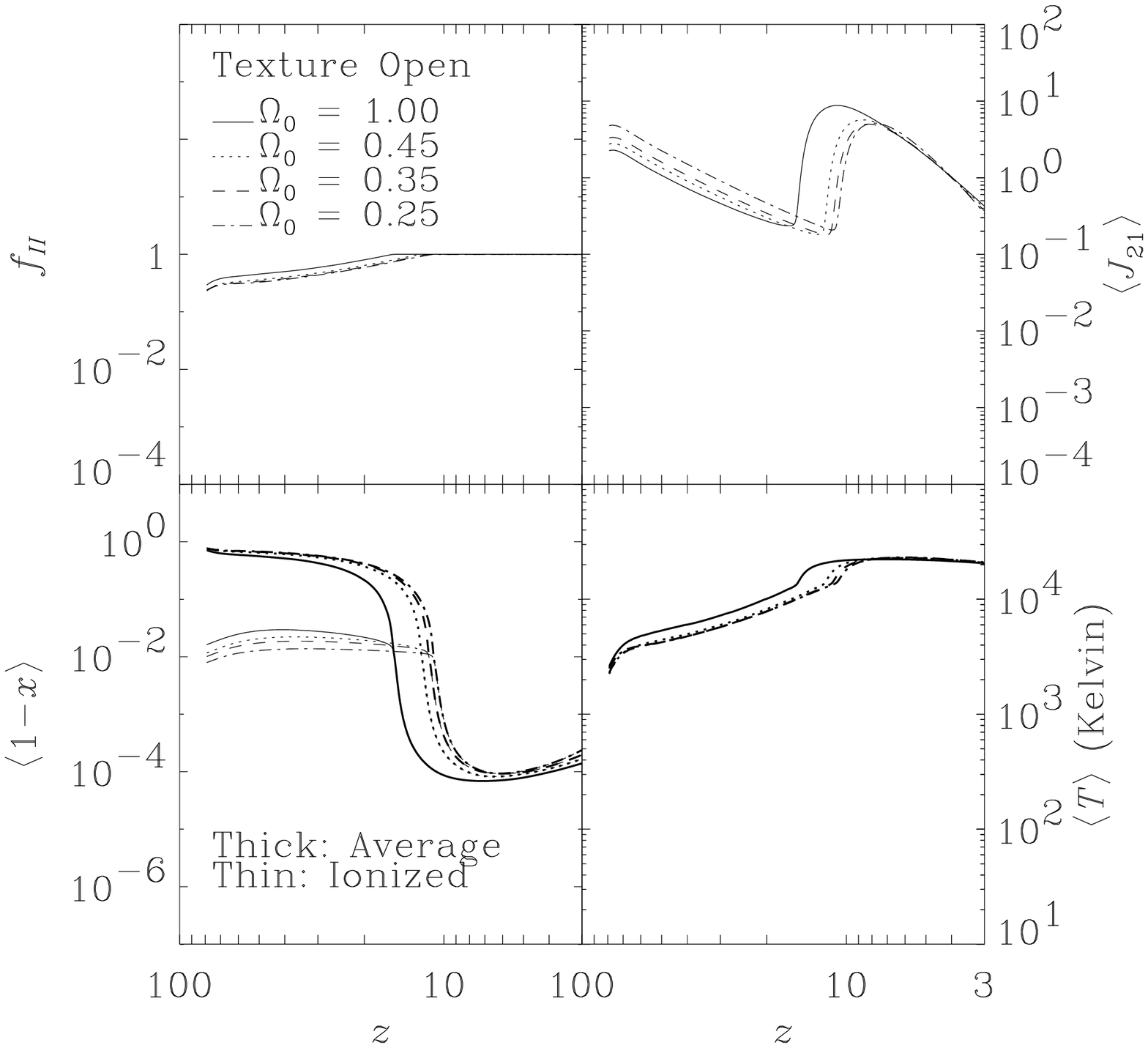]
{
Results of reionization calculation for open texture models.
\label{fig:text_open_fourplot}
}

\figcaption[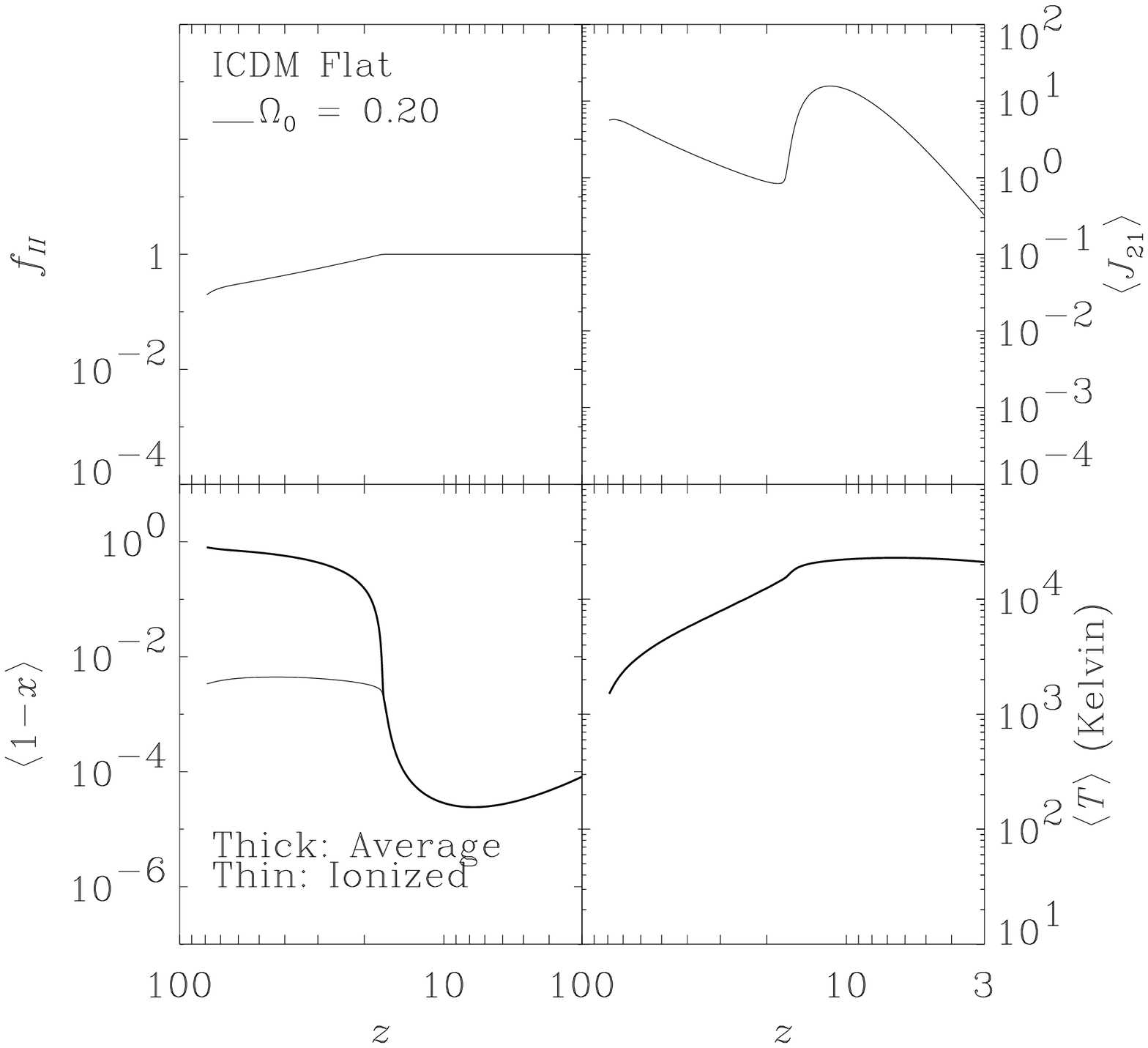]
{
Results of reionization calculation for ICDM model.
\label{fig:icdm_flat_fourplot}
}

\figcaption[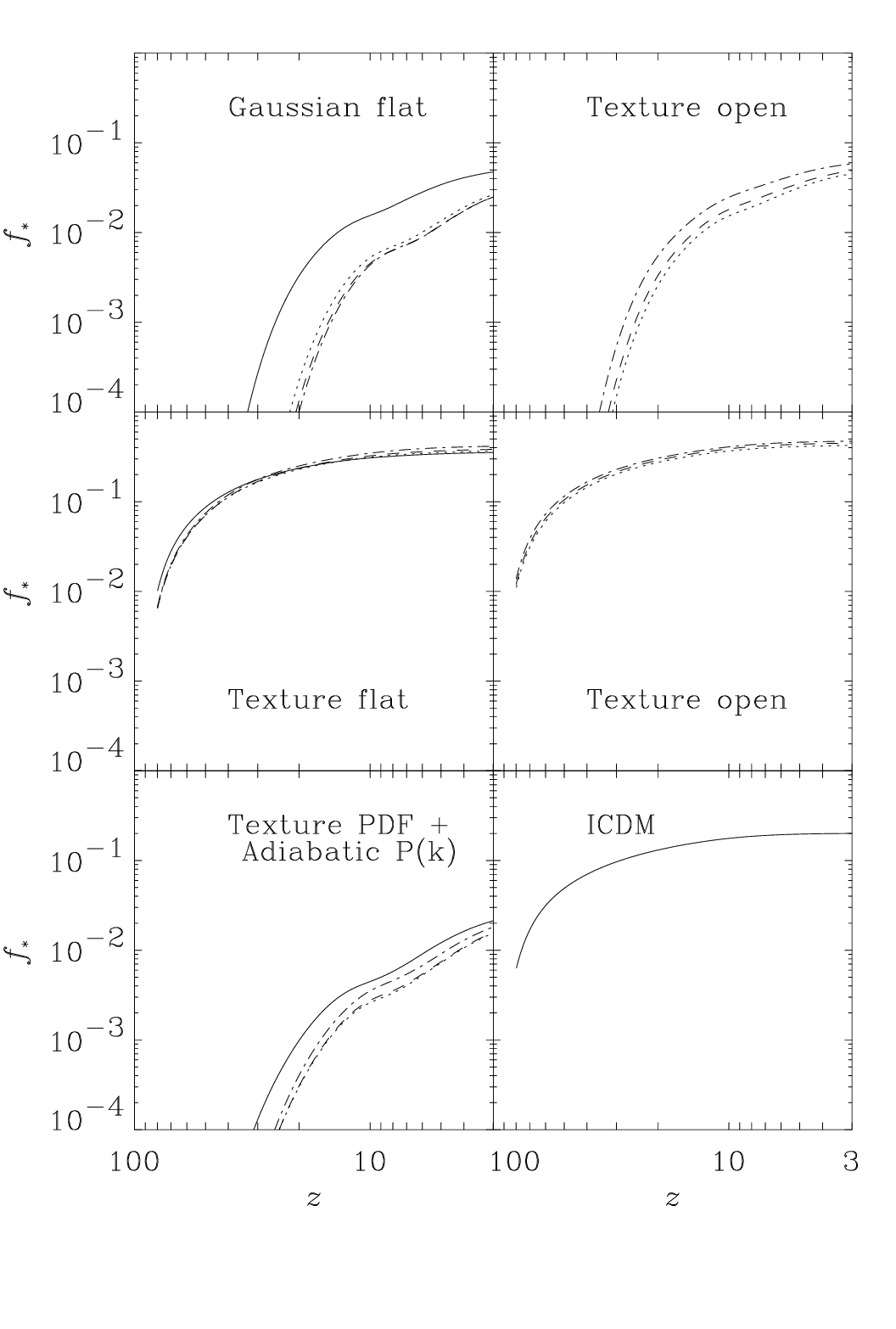]
{
Fraction of baryons in stars/quasars for different models.  
Except for the ICDM model, the different
line styles denote $\Omega_0=1$ (solid), 0.45 (dotted), 0.35 (dashed), 
and 0.25.  For the ICDM model, which 
only has one background cosmology, $\Omega_0=0.2$.  
\label{fig:fstar}
}

\figcaption[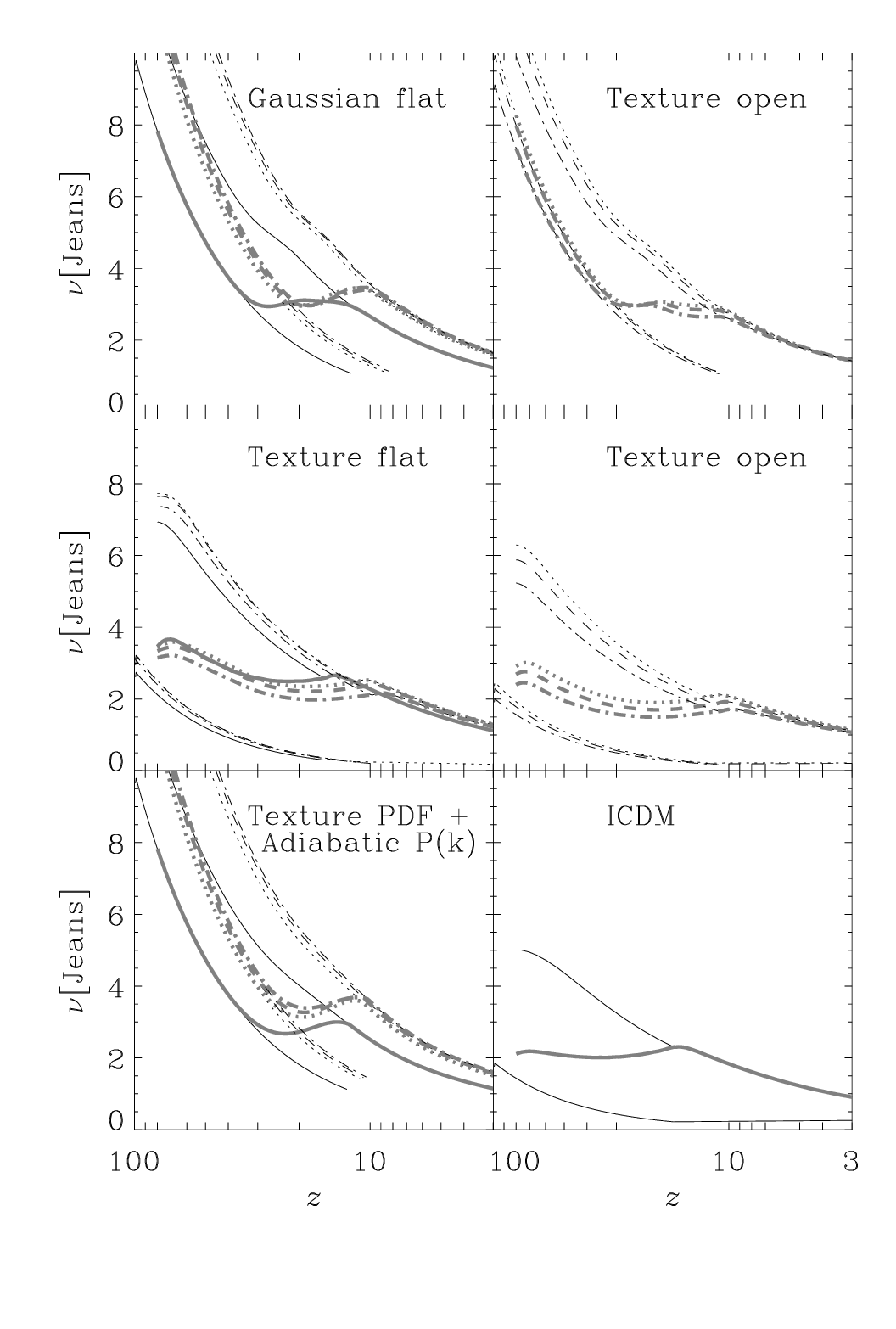]
{
Degree of nonlinearity $\nu$ of Jeans scale for 
different models.  The correspondence between line styles
and $\Omega_0$ are the same as in Figure \ref{fig:fstar}.
The upper thin lines are $\nu$ for the ionized phase,
and lower thin lines are $\nu$ for the neutral phase, and the
thick lines in between are the average values of $\nu$.
\label{fig:nuplot}
}

\figcaption[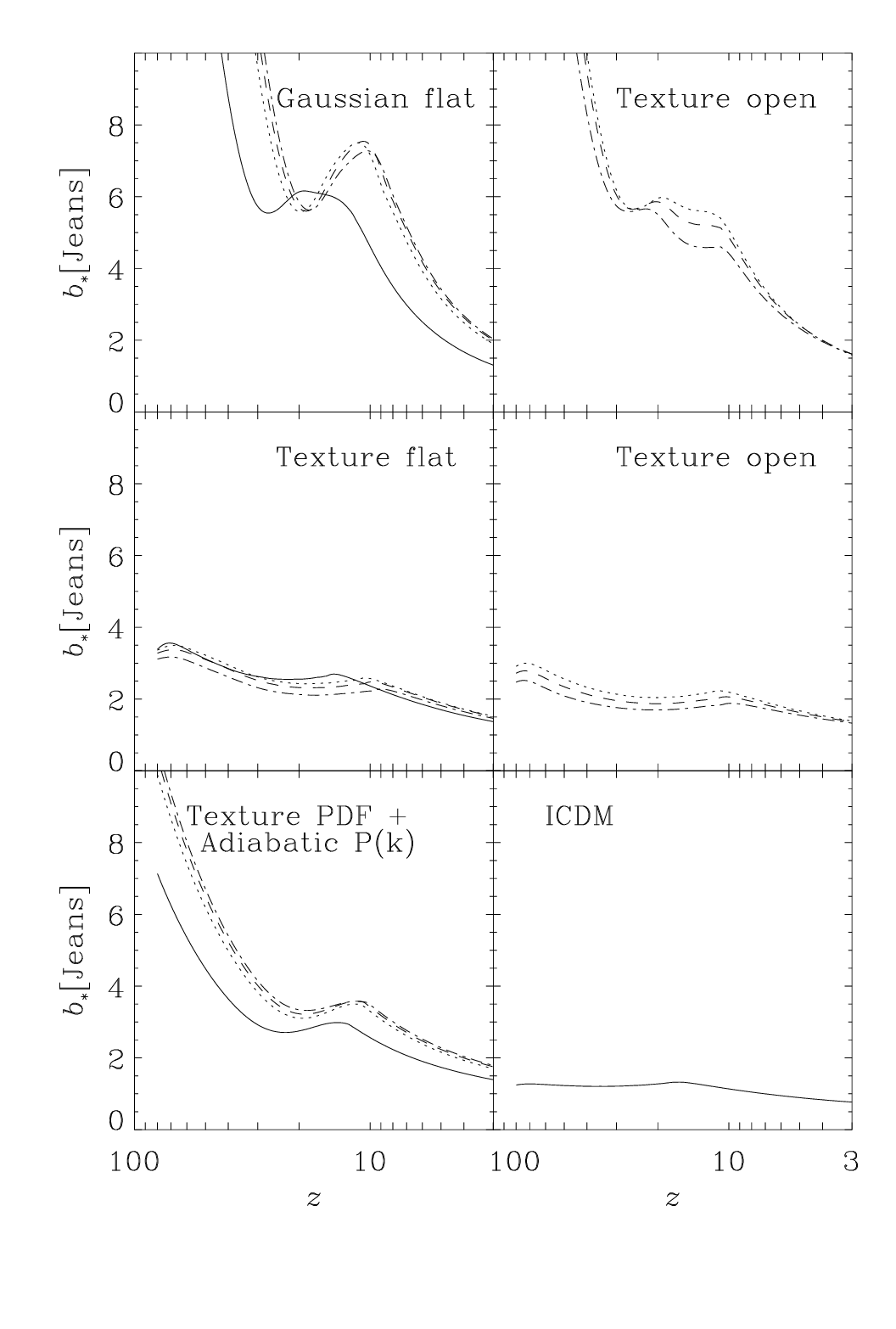]
{
Bias at birth $b_\ast$ of Jeans scale objects for
different models.  The definitions of line styles are
the same as in Figure \ref{fig:nuplot}.
\label{fig:bplot}
}

\figcaption[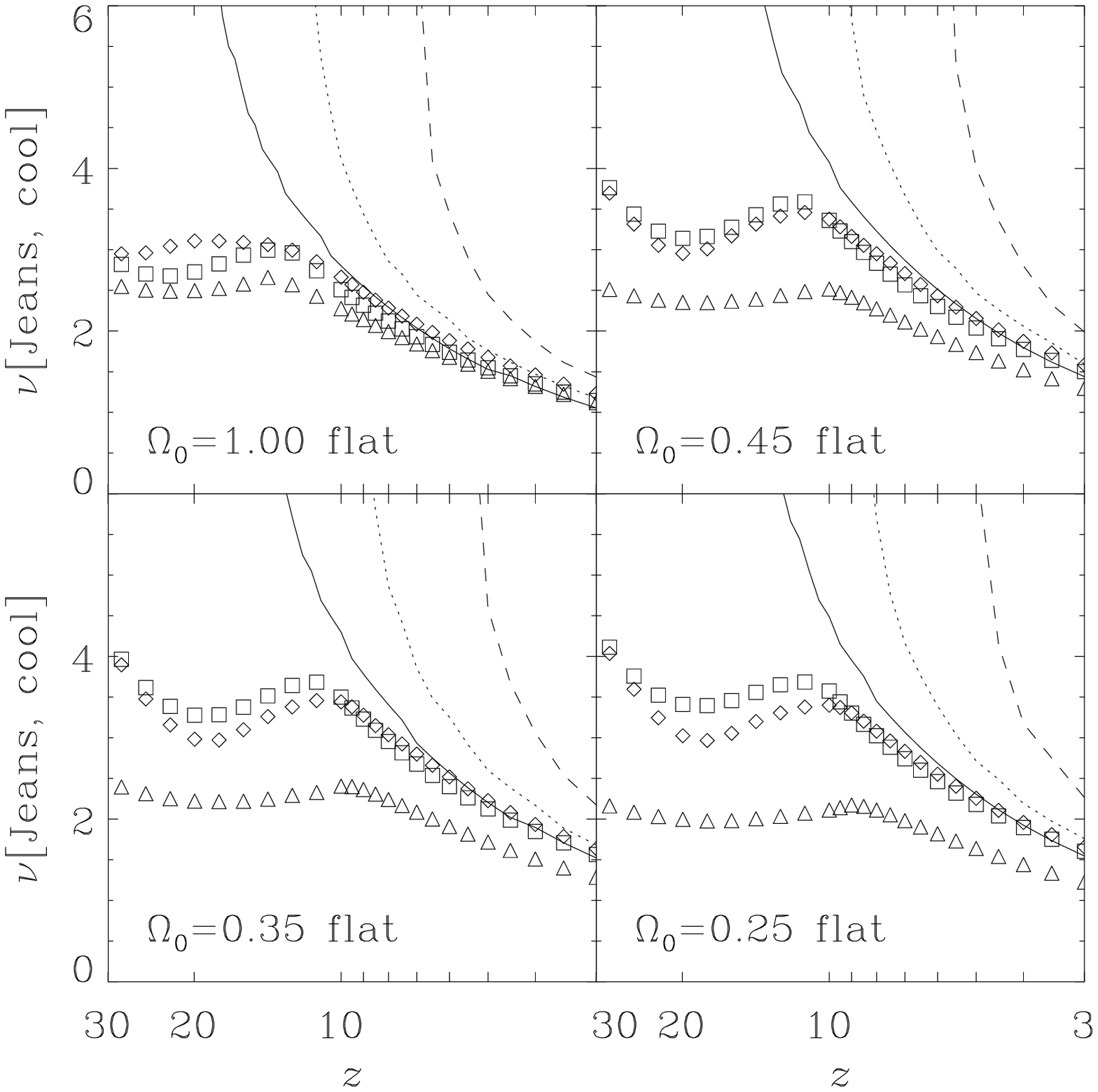]
{
Limits on $\nu$ from Jeans criterion and cooling 
efficiency 
for flat Gaussian and texture Models.  The adiabatic Gaussian G+ACDM model
is denoted by diamonds $\Diamond$, the texture model T+CDM is denoted
by triangles $\bigtriangleup$, and the texture + adiabatic model
T+ACDM is denoted by squares $\Box$.  
The contour lines indicate the value of
$\nu_c(z,\epsmin)$ with the minimum fractions of the
mass of the halo can cool in a dynamical time having
values of $\epsmin = 10^{0,-0.5,-1}$ 
(solid, dotted, dashed lines).
\label{fig:nu_j_cool_comp_flat}
}

\figcaption[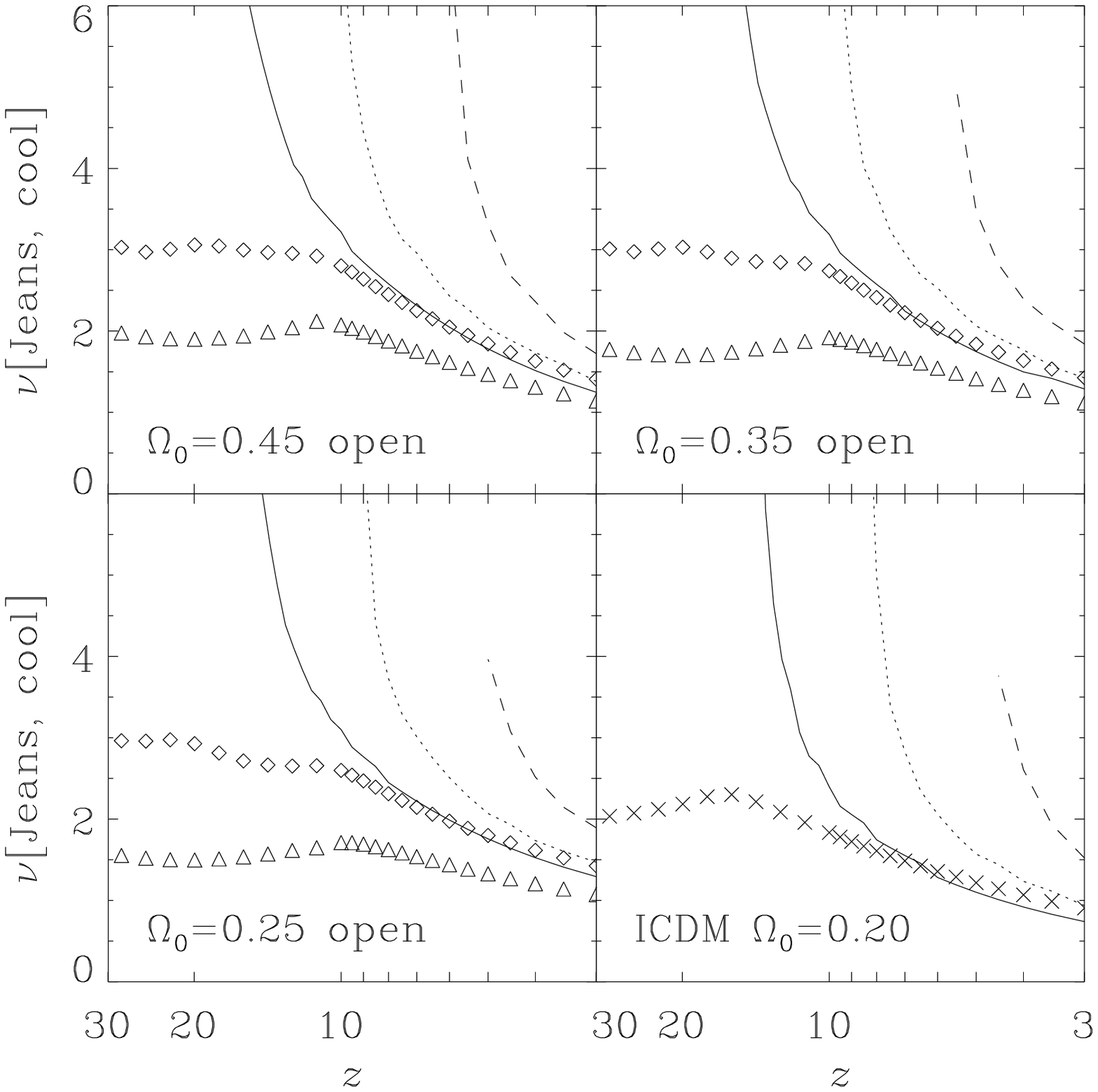]
{
Limits on $\nu$ from Jeans criterion and cooling efficiency 
for open Gaussian and texture models, and ICDM model. 
The adiabatic Gaussian model G+ACDM 
is denoted by diamonds $\Diamond$, the texture model T+CDM is denoted
by triangles $\bigtriangleup$, and the ICDM model is denoted by
the crosses $\times$.  The contour lines indicate the value of
$\nu_c(z,\epsmin)$ with the minimum fractions of the
mass of the halo can cool in a dynamical time having
values of $\epsmin = 10^{0,-0.5,-1}$ 
(solid, dotted, dashed lines).
\label{fig:nu_j_cool_comp_open_icdm}
}

\clearpage

\figcaption[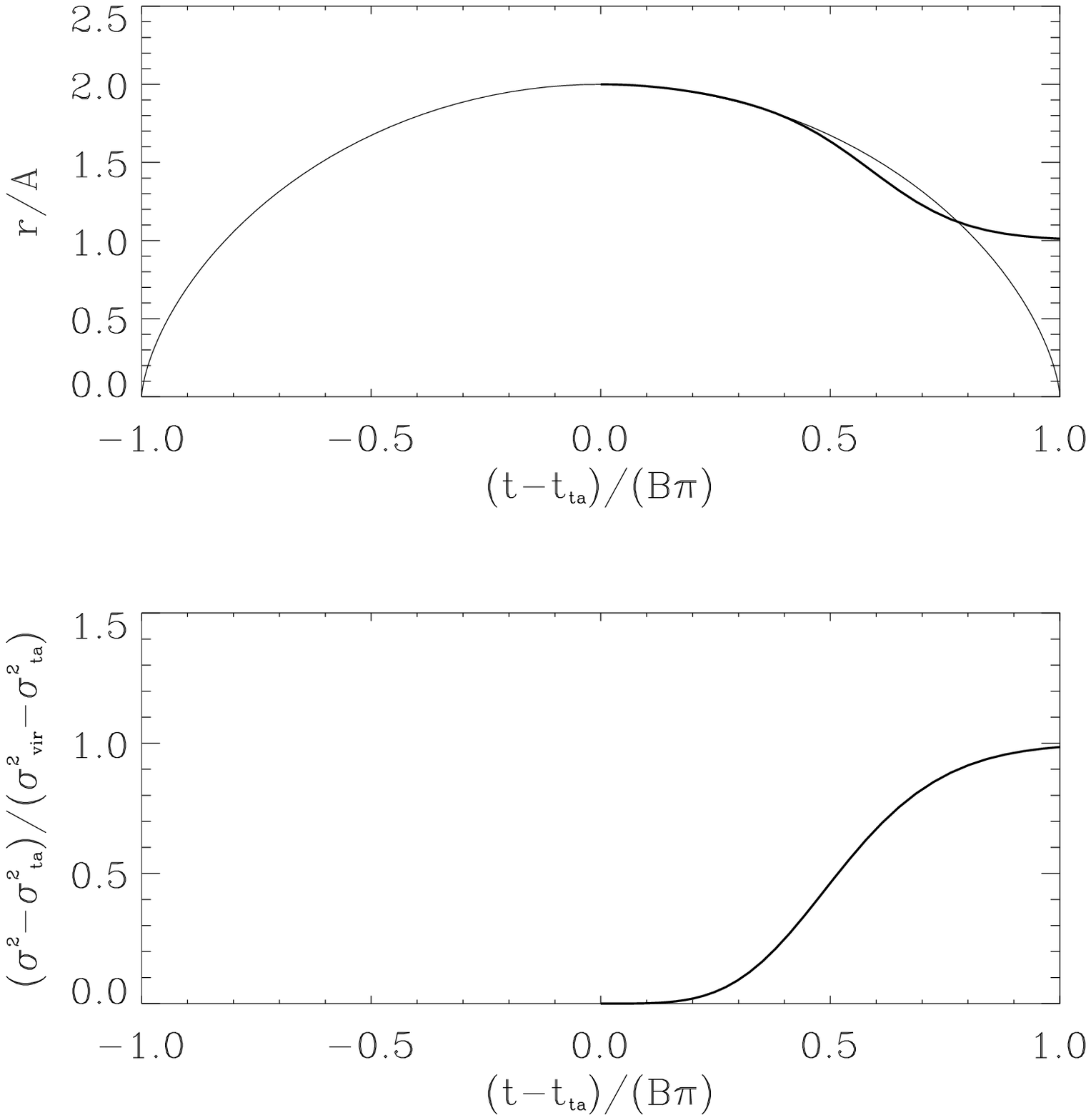]
{
Top figure gives the radius as a function of time
for uniform spherical collapse (thin line) and our
functional form for collapse and virialization with
virial radius equal to half the
maximum radius (thick line).  The lower figure gives the
function form of the heating of gas from its velocity
dispersion at turnaround $\sigma^2_{\rm ta}$ to the 
virial equilibrium with velocity dispersion $\sigma^2_{\rm vir}$.
\label{fig:sphcolfunc}
}

\figcaption[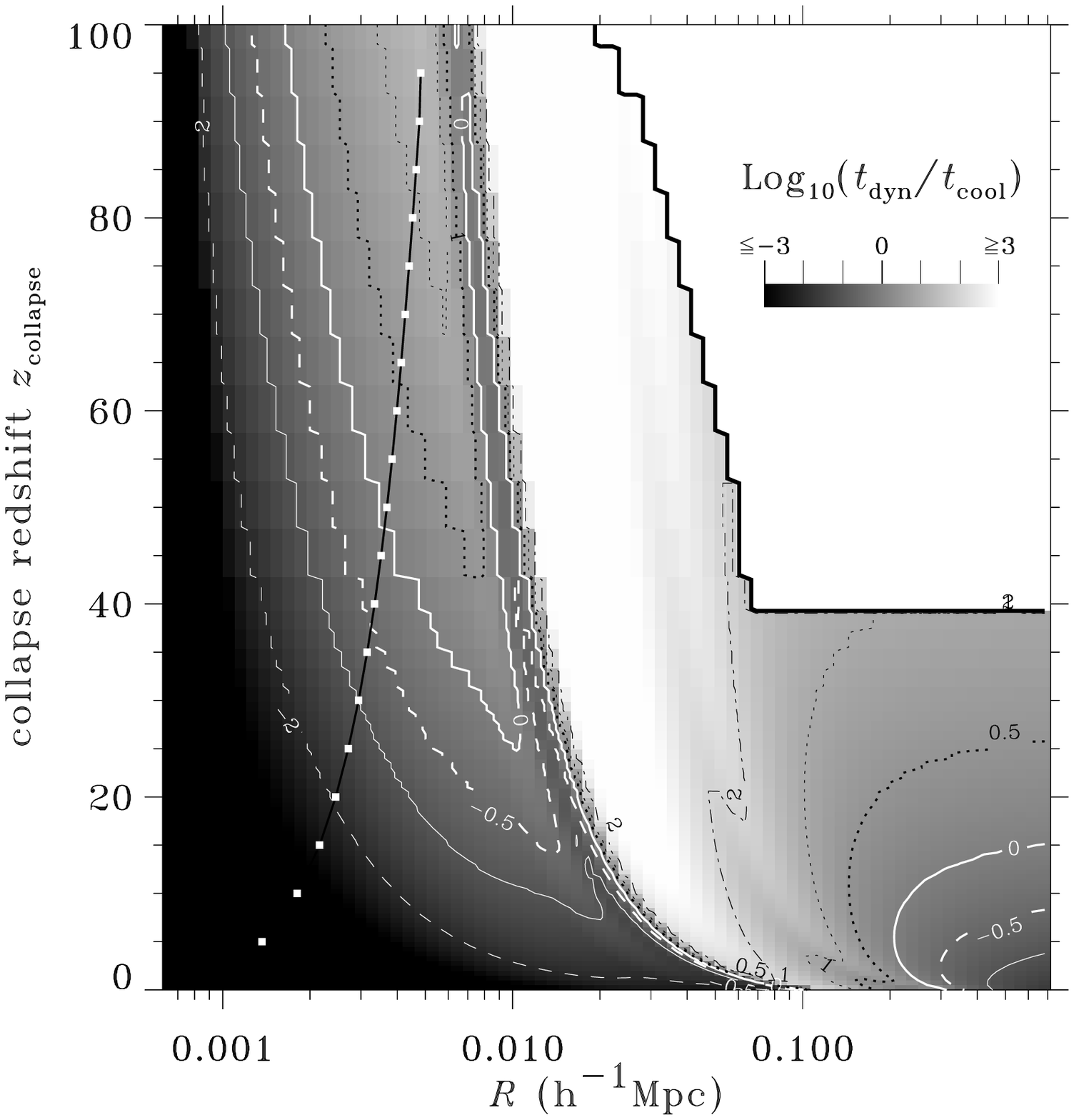]
{
The ratio of dynamical time to cooling time for $\Omega_0=1$, as
a function of collapse redshift and comoving radius.  Also shown
is the comoving Jeans length of the neutral phase (white boxes).  
The contours mark 
where $\tdyn/\tcool = 10^{-2,-1,-0.5, 0, 0.5, 1, 2}$.  
\label{fig:coolgrid_Omega_one}
}

\figcaption[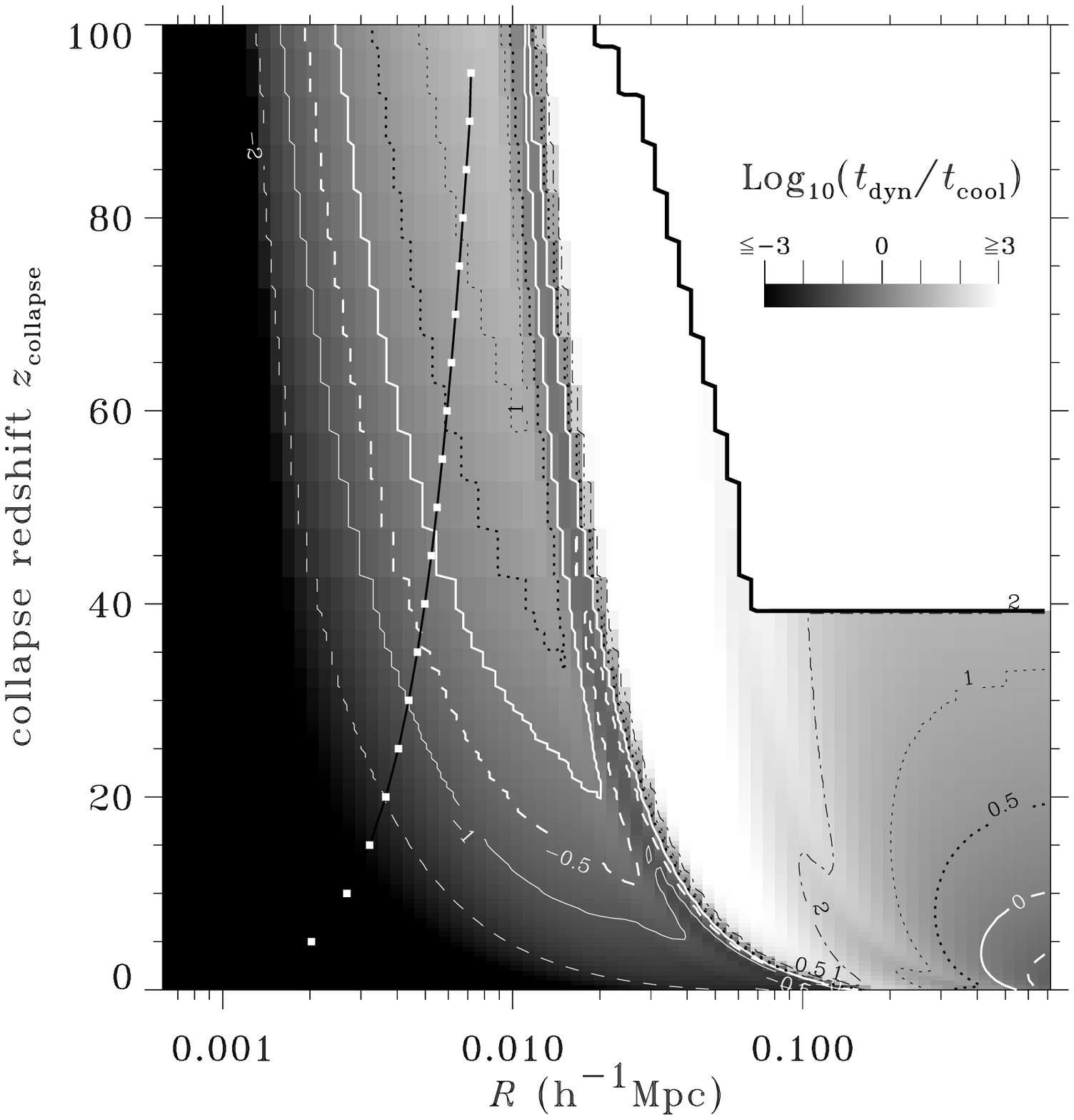]
{
The ratio of dynamical time to cooling time for $\Omega_0=0.35$, as
a function of collapse redshift and comoving radius.  
The notation is the same as in Figure~\ref{fig:coolgrid_Omega_one}
\label{fig:coolgrid_Omega_zerothreefive}
}

\clearpage 

\begin{deluxetable}{cccccccccc}
\tablecolumns{10}
\tablewidth{0pc}
\tablecaption{Cosmological Models Considered and 
	Calculated Results Related to the Thermal and
	Ionization History of the Universe
	\label{tab:results}}
\tablehead{
\multicolumn{3}{c}{Cosmological} & 
	\multicolumn{7}{c}{Results}
\\
\cline{4-10}
\multicolumn{3}{c}{Parameters} & Reheating &
	\multicolumn{2}{c}{Reionization}  &
	\multicolumn{2}{c}{CMBR} &
	\multicolumn{2}{c}{IGM} 
\\
\hline
\colhead{$\Omega_0$} & \colhead{$\Omega_{\Lambda}$} & \colhead{$\sigma_8$} & 
	early &
	\multicolumn{2}{c}{$z$ for $\langle 1-x\rangle$} of &
	\colhead{$y$} & \colhead{$\tau_{\rm rec}$} &
	\colhead{$\tau_{\rm GP}$} & \colhead{$f_\ast$}
\\
\colhead{} & \colhead{} & \colhead{} &
	\colhead{or late} &
	\colhead{\phn $10^{-1}$} & 
		\colhead{\phn $10^{-3}$} &
	\colhead{$(\times 10^{-7})$} & \colhead{} &
	\multicolumn{2}{c}{(at $z=3$)}
}
\startdata
\sidehead{Gaussian Adiabatic CDM}
1.0\phn & 0.0\phn & 0.60
	& late  
	& 13.1 & 10.3 
	& $1.4$  & 0.069 
	&  0.14 & 0.047
\\
0.45 & 0.55 & 0.71 
	& late  
	& 10.6 & \phn 7.6  
	& $1.3$  & 0.065
	& 0.17  & 0.027
\\
0.35 & 0.65 & 0.76 
	& late  
	& 10.2 &\phn 7.3 
	& $1.4$  & 0.068 
	& 0.17  & 0.025
\\
0.25 & 0.75 & 0.85  
	& late  
	&\phn 9.8 &\phn 7.2 
	& $1.6$  & 0.077
	& 0.19  & 0.025
\\
0.45 & 0.0\phn & 0.68 
	& late  
	&  11.9 &\phn 9.7 
	& $1.8$  & 0.087 
	& 0.17  & 0.045
\\
0.35 & 0.0\phn & 0.71 
	& late  
	& 11.8 &\phn 9.8
	& $2.1$  & 0.099
	& 0.17  & 0.049
\\
0.25 & 0.0\phn & 0.76 
	& late  
	& 11.7 & 10.0
	& $2.6$  & 0.12\phn
	& 0.21  & 0.058
\\
\\[-2pc]
\sidehead{Textures + CDM}
1.0\phn & 0.0\phn & 0.37 
	& early
	&  17.2 & 14.4
	& $5.8$  & 0.28\phn
	&  0.08 & 0.35\phn
\\
0.45 & 0.55 & 0.50 
	& early
	& 12.8 & 10.6
	& $6.5$  & 0.32\phn
	& 0.13  & 0.37\phn
\\
0.35 & 0.65 & 0.56 
	& early
	&  12.0 & 10.0
	& $6.6$  & 0.35\phn
	& 0.14  & 0.38\phn
\\
0.25 & 0.75 & 0.66 
	& early
	& 11.0 &\phn 9.1
	& $3.9$ & 0.38\phn
	& 0.16  & 0.42\phn
\\
0.45 & 0.0\phn & 0.47 
	& early
	&  13.6 & 11.3
	& $5.9$  & 0.33\phn
	& 0.11  & 0.42\phn
\\
0.35 & 0.0\phn & 0.50 
	& early
	& 12.7 & 10.6
	& $4.6$  & 0.43\phn
	& 0.11  & 0.45\phn
\\
0.25 & 0.0\phn & 0.56 
	& early
	& 12.3  & 10.1
	&  $1.8$ & 0.46\phn
	& 0.12  & 0.48\phn
\\
\\[-2pc]
\sidehead{Textures + Adiabtic CDM power spectrum}
1.0\phn & 0.0\phn & 0.35 
	& late
	& 13.6 & 10.8
	& $1.0$  & 0.065
	& 0.15 & 0.021 
\\
0.45 & 0.55 & 0.47 
	& late
	& 11.8 &\phn 9.7 
	& $1.2$  & 0.073
	& 0.18  & 0.016
\\
0.35 & 0.65 & 0.53 
	& late
	& 11.6  &\phn 9.6
	& $1.3$  & 0.074
	& 0.19  & 0.016
\\
0.25 & 0.75 & 0.62 
	& late
	& 11.2 &\phn 9.5
	& $1.7$ & 0.097
	& 0.22 &  0.018
\\
\\[-2pc]
\sidehead{Non-Gaussian Isocurvature CDM}
0.20 & 0.80 & 0.90 
	& early
	& 18.9 & 16.5 
	& 11\phd  & 0.53\phn
	& 0.44 & 0.20\phn
\\
\enddata
\end{deluxetable}

\clearpage

\begin{deluxetable}{cl}
\tablecolumns{2}
\tablewidth{0pc}
\tablecaption{Chemical Reactions Involving $\htwo$, $\htwop$, and $\hm$
\label{tab:h2reacts}
}
\tablehead{
\colhead{Reaction number} & \colhead{Reaction} 
}
\startdata
 (7) &  $\hn   +  e^{-}  \rightarrow
                                \hm     +    \gamma$
  \\
 (8) &  $\hm    +  \hn \rightarrow
                                \htwo    +    e^{-}$
  \\
 (9) &  $\hn    +  \hp\rightarrow
                                \htwop + \gamma$
  \\
 (10) &  $\htwop + \hn\rightarrow
                                 \htwo   +    \hp$
  \\
 (11) &  $\htwo  +  \hp\rightarrow
                                 \htwop +    \hn$
  \\
 (12) &  $\htwo  +  e^{-}\rightarrow
                                 2\hn   +    e^{-}$
  \\
 (13) &  $\htwo  +  \hn\rightarrow
                                 3\hn  $
  \\
 (14) &  $\hm  +   e^{-} \rightarrow
                                 \hn     +   2e^{-}$
  \\
 (15) &  $\hm  +  \hn   \rightarrow
                                 2\hn    +    e^{-}$
  \\
 (16) &  $\hm +  \hp \rightarrow
                                 2\hn$
  \\
 (17) &  $\hm  +  \hp \rightarrow
                                 \htwop +   e^{-}$
  \\
 (18) &  $\htwop +  e^{-}\rightarrow
                                 2\hn$  
  \\
 (19) &  $\htwop +  \hm\rightarrow
                                 \htwo   +   \hn$
  \\
 (23) &  $\hm   +  \gamma \rightarrow
                                 \hn    +    e^{-}$
  \\
 (24) &  $\htwo   +  \gamma\rightarrow
                                 \htwop  +   e^{-}$
  \\
 (25) &  $\htwop +  \gamma\rightarrow
                                 \hn    +   \hp$
  \\
 (26) &  $\htwop  +  \gamma
                                \rightarrow
                                 2\hp +   e^{-}$
  \\
 (27) &  $\htwo   +   \gamma
                                \rightarrow  {\rm H}_2^* \rightarrow
                                 2\hn$
  \\
 (28) &  $\htwo   +   \gamma
                                 \rightarrow
                                 2\hn$    
  \\
\enddata
\end{deluxetable}

\input{figs}

\end{document}

%% file: figs.tex
\begin{figure}
\figurenum{1}
\resizebox{\textwidth}{!}{\includegraphics{fig1.eps}}
\caption{}
\end{figure}
\clearpage

\begin{figure}
\figurenum{2}
\resizebox{\textwidth}{!}{\includegraphics{fig2.eps}}
\caption{}
\end{figure}
\clearpage

\begin{figure}
\figurenum{3}
\resizebox{\textwidth}{!}{\includegraphics{fig3.eps}}
\caption{}
\end{figure}
\clearpage

\begin{figure}
\figurenum{4}
\resizebox{\textwidth}{!}{\includegraphics{fig4.eps}}
\caption{}
\end{figure}
\clearpage

\begin{figure}
\figurenum{5}
\resizebox{\textwidth}{!}{\includegraphics{fig5.eps}}
\caption{}
\end{figure}
\clearpage

\begin{figure}
\figurenum{6}
\resizebox{\textwidth}{!}{\includegraphics{fig6.eps}}
\caption{}
\end{figure}
\clearpage

\begin{figure}
\figurenum{7}
\resizebox{\textwidth}{!}{\includegraphics{fig7.eps}}
\caption{}
\end{figure}
\clearpage

\begin{figure}
\figurenum{8}
\resizebox{\textwidth}{!}{\includegraphics{fig8.eps}}
\caption{}
\end{figure}
\clearpage

\begin{figure}
\figurenum{9}
\resizebox{\textwidth}{!}{\includegraphics{fig9.eps}}
\caption{}
\end{figure}
\clearpage

\begin{figure}
\figurenum{10}
\resizebox{!}{7in}{\includegraphics{fig10.eps}}
\caption{}
\end{figure}
\clearpage

\begin{figure}
\figurenum{11}
\resizebox{!}{7in}{\includegraphics{fig11.eps}}
\caption{}
\end{figure}
\clearpage

\begin{figure}
\figurenum{12}
\resizebox{!}{7in}{\includegraphics{fig12.eps}}
\caption{}
\end{figure}
\clearpage

\begin{figure}
\figurenum{13}
\resizebox{\textwidth}{!}{\includegraphics{fig13.eps}}
\caption{}
\end{figure}
\clearpage

\begin{figure}
\figurenum{14}
\resizebox{\textwidth}{!}{\includegraphics{fig14.eps}}
\caption{}
\end{figure}
\clearpage

\begin{figure}
\figurenum{15}
\resizebox{\textwidth}{!}{\includegraphics{fig15.eps}}
\caption{}
\end{figure}
\clearpage

\begin{figure}
\figurenum{16}
\resizebox{\textwidth}{!}{\includegraphics{fig16.eps}}
\caption{}
\end{figure}
\clearpage

\begin{figure}
\figurenum{17}
\resizebox{\textwidth}{!}{\includegraphics{fig17.eps}}
\caption{}
\end{figure}
\clearpage